\documentclass[sort&compress]{elsarticle}
\pdfoutput=1

\usepackage{multirow}
\usepackage{amsmath}
\usepackage{bm}
\usepackage{booktabs}
\usepackage{subfloat}

\usepackage{hyperref}
\usepackage{subfig}
\usepackage{algorithm}
\usepackage[table]{xcolor}
\definecolor{blueish}{rgb}{0.8,0.8,1.0}

\usepackage{verbatim}
\usepackage{algorithmic} 
\newtheorem{lemma}{Lemma}
\newtheorem{theorem}{Theorem}

 \newtheorem{proposition}{Proposition}
 
 \usepackage{url}
 \usepackage{algorithm}

 \usepackage{algorithmic}
 \usepackage{graphicx}
\graphicspath{{scripts/}{plots/}{.}{../BitmapIndexCpp/dbash/}{../BitmapIndexCpp/bash/}{../BitmapIndexCpp/dbash/RLEmodelling/}{ExperimentalSoftware/}}
\newtheorem{definition}{Definition}

\usepackage{color}

\newcommand{\daniel}[1]{\textbf{{\small{\color{magenta}DL}: #1{\color{magenta}$\circ$}}}}
\newcommand{\owen}[1]{\textbf{{\small{\color{red}OK}: #1{\color{red}$\circ$}}}}

\newcommand{\cut}[1]{}

\usepackage{amssymb}
\newenvironment{proof}{\textbf{Proof.}}{ $\square$ }

\usepackage{ifpdf}
\ifpdf
\else

\fi

\begin{document}

\begin{frontmatter}




\title{Reordering Columns for Smaller Indexes} 


 \author[UQAM]{Daniel Lemire\corref{cor1}} \ead{lemire@acm.org} 
 \author[UNB]{Owen Kaser} \ead{o.kaser@computer.org}  

 \address[UQAM]{\scriptsize LICEF, Universit\'e du Qu\'ebec \`a Montr\'eal (UQAM), 100 Sherbrooke West, Montreal, QC, H2X 3P2 Canada
}

 \address[UNB]{\scriptsize Dept.\ of CSAS, University of New Brunswick, 100 Tucker Park Road, Saint John, NB, Canada}
 \cortext[cor1]{Corresponding author. Tel.: 00+1+514 987-3000 ext. 2835; fax: 00+1+514 843-2160.}








\begin{abstract}
Column-oriented indexes---such as 
projection 
 or bitmap indexes---are compressed by 
run-length encoding 
to reduce  storage and increase speed.
Sorting the tables improves compression.
On realistic data sets, permuting the columns in the right order before sorting
can reduce the number of runs by a factor of two or more.
Unfortunately, determining the best column order is NP-hard.
For many cases,
we prove that the number of runs in table columns is
minimized if  we sort columns by increasing cardinality.
Experimentally, 
sorting based on Hilbert space-filling
curves is poor at minimizing the number of runs.
\end{abstract}

\end{frontmatter}

\section{Introduction}
\label{sec:intro}

Many database queries have low selectivity.
In these instances, we may need to  load the content
of entire columns. 
To improve performance and reduce memory usage, we compress columns with  lightweight 
techniques such as  run-length encoding (RLE). Yet  RLE compression is better  
if there are  long runs of identical values within columns. 

Meanwhile, sorting reduces the number of these \emph{column runs}. 
In fact, sorting the table before indexing can improve the speed of an index by nearly a factor of ten~\cite{arxiv:0901.3751},
while reducing the memory usage in a comparable manner.

Yet there are many ways to  sort a table, and 
we are motivated to sort the table in the best possible manner.
Adabi et al.\ recommend lexicographic sorting with ``low cardinality columns serv[ing] as the leftmost sort orders''~\cite{1142548}.
We want to justify  this empirical recommendation. 

For uniformly distributed tables, we show that
 sorting lexicographically
with the columns in increasing cardinality is asymptotically optimal---for large column cardinalities. 
Furthermore, we show how to extend this result to all column cardinalities.
As an additional contribution, we  bound the suboptimality of 
sorting lexicographically 
for the problem of minimizing the number of runs.
With this analytical bound, we show that for several 
 realistic tables, sorting  is
3-optimal or better as long as the 
columns are ordered in increasing cardinality.


We present our results in 
four 
steps: modeling (\S~\ref{sec:modellingrle}),
a priori bounds (\S~\ref{sec:yesitisnphard} and \S~\ref{sec:lex-and-gc-sorting}), analysis of synthetic cases (\S~\ref{sec:Increasing-cardinality-order}) and experiments (\S~\ref{sec:experiments}).
Specifically, the paper is organized as follows:
\begin{itemize}
\item There are many possible RLE implementations. In \S~\ref{sec:modellingrle}, we 
propose  to count  column runs as a simplified cost model.
\item In \S~\ref{sec:yesitisnphard}, we prove that minimizing the number of runs by
row reordering is NP-hard.
\item 
 In \S~\ref{sec:lex-and-gc-sorting},
we review several orders used to sort tables in databases~: the lexicographical order,
the reflected Gray-code order, and so on. 
We regroup many of these orders into a  family: the recursive orders.
In \S~\ref{sec:significanceofcolumnorder}, 
we bound  the suboptimality
of sorting as a heuristic
  to minimize the number of runs. 
In \S~\ref{sec:colr}, we prove that determining the best column order is NP-hard.
\item In \S~\ref{sec:Increasing-cardinality-order}, we analytically determine the best column
order for some synthetic cases. Specifically, in \S~\ref{sec:complete-tables}, we analyze
 tables where all possible tuples are present. In \S~\ref{sec:uniformlydistributed},
we consider the more difficult problem of uniformly distributed tables. We 
first prove that for high cardinality columns, organizing the columns in increasing cardinality
is best at minimizing the number of runs (see Theorem~\ref{theorem:genericresult}). In \S~\ref{sec:lowcardlexico} and
\S~\ref{sec:ExpectednumberofrunsafterGraycodesorting}, we show how to extend this result
to low cardinality columns for the lexicographical and reflected Gray-code orders.
\item 
Finally, we experimentally verify the importance of column ordering in \S~\ref{sec:experiments},
and assess other factors such as column dependencies. We show 
that an order based on Hilbert space-filling curves~\cite{hamilton2007chi} is not competitive
to minimize the number of runs.
\end{itemize}

\section{Modeling RLE compression by the number of column runs}
\label{sec:modellingrle}

RLE  
compresses 
 long runs of
identical values: it replaces any run by the number
of repetitions followed by the value being repeated. For example, 
the sequence  11111000 becomes 5--1, 3--0.  
In column-oriented databases,
RLE makes many queries faster:
sum, average, median, percentile, and arithmetic
operations over several columns~\cite{253268}.


There are many variations on RLE:
\begin{itemize}
\item Counter values can be stored using fixed-length counters.
In this case, any run whose length exceeds the capacity of the counter 
is stored as multiple runs. 
For example,
Adabi et al.~\cite{1142548} use a fixed number of bits for the tuple's value, start position, and
run length. 
We can also use variable-length counters~\cite{scholer2002cii,323905,zobel2006inverted,yan2009inverted,db2luw2009,1034897,1367550,moffat2000binary} or quantized codes~\cite{1247525}. 
\item 
When values are represented using fewer bits than the
counter values, we may add the following convention: a counter
is only present after the same value is repeated twice.
\item In the same spirit, we may use a single bit to indicate whether 
a counter follows the current value. This is convenient
if we are transmitting 7-bit ASCII characters using 8-bit words~\cite{10.1109/TSE.1985.231852,672970}.
\item It might be inefficient to store
short runs using value-counter pair.
Hence, we may leave short runs uncompressed 
 (BBC~\cite{874730}, WAH~\cite{wu2006obi} or EWAH~\cite{arxiv:0901.3751}). 
\item Both the values and the counters have some statistical
distributions. If we know these distributions, more efficient
encodings are possible by combining statistical compression
with RLE---such as Golomb coding~\cite{golomb1966rle}, Lempel-Ziv, Huffman, or arithmetic encoding. Moreover, if we expect the values to appear in some 
specific order, we can store a delta instead of the value~\cite{holloway2008rod}.
For example, the list of values 00011122, can be coded as 
the (diffed-values,counter) pairs $(1,3) (1,3), (1,2)$. This can be used to
enhance compression further.
\item To support binary search 
within an RLE array,
we may store not only the value and the repetition count, but also
the location of the run
~\cite{10.1109/TSE.1985.231852,1321546,moffat1996sii}, or we may use
a B-tree~\cite{1353407}.
\item Instead of compressing the values themselves, we may compress their
bits.  In  bitmap indexes,
for any given column, several bitmaps can be individually compressed by RLE.
\end{itemize}

It would be futile to attempt to analyze mathematically all possible 
applications of 
RLE to database indexes. Instead, we count  runs of identical values.
That is, if 
$r_i$ \label{ri-defined} is the number of runs in column $i$ and there are $c$~columns, we compute $\sum_{i=1}^c r_i$\footnote{
A table of notation can be found
in Appendix~\ref{sec:notation}.} (henceforth \textsc{RunCount}). 

\section{Minimizing the number of runs by row reordering is NP-hard}
\label{sec:yesitisnphard}

We want to minimize \textsc{RunCount} by row reordering.
Consider a related problem over Boolean matrices~\cite{johnson2004clb}: minimizing the number of runs \textbf{of ones} in rows  by column reordering.
This ``Consecutive Block Minimization'' problem (CBMP) is NP-hard
\cite[SR17]{gare:gandj},\cite{blockminisnphard}\footnote
{Another NP-hardness proof was later given by Pinar and Heath~\cite{331562}.}.
Yet, even if we transpose the matrix, CBMP is not equivalent to the \textsc{RunCount} minimization problem.
Indeed, 
both sequences 001100 and 000011 have a single run of ones. Yet the sequence
001100 has three runs whereas the second sequence (000011)
has only two runs. Moreover, the \textsc{RunCount} minimization problem is not limited to binary data.
To our knowledge, there is no published proof that minimizing \textsc{RunCount} by row reordering is NP-hard.
Hence, we provide the following result.

\begin{lemma}\label{lemma:runcountnphard}
Minimizing \textsc{RunCount} by row reordering is NP-hard.
\end{lemma}
\begin{proof}
We prove the result by reduction from the Hamiltonian path problem, which remains NP-hard even
if a starting and ending vertex are specified~[GT39]\cite{gare:gandj}.
Consider any connected graph $G$ having $n$~vertices and $m$~edges, and let $s$ and $t$ be respectively the
beginning and end of the required Hamiltonian path.

Consider the incidence matrix of such a graph. There is a row for each vertex, and
a column for each edge. The value of the matrix is one if the edge connects with the vertex, and zero
otherwise.  Each column has only two ones; thus it has either
\begin{enumerate}
\item two runs (if the ones are consecutive, and either at the top or bottom of the column)
\item three runs (if the ones are consecutive but not at the top or bottom of the column, or if there are ones
at the top and bottom)
\item four runs (if the ones are not consecutive, but a one is at the top or at the bottom), or
\item five runs (in all other cases).
\end{enumerate}
Thus, the number of column runs in this incidence matrix is less than $5m$.


We modify the incidence matrix by adding $10m$~new columns. These columns contain only zeros,
except that $5m$~columns have the value one on the row corresponding to vertex $s$, and
$5m$~other columns have the value one on the row corresponding to vertex $t$ (see Fig.~\ref{fig:proof1}).
These new columns have either 2~runs or 3~runs depending on whether the rows corresponding
to $s$ and $t$ are first, last or neither.

\begin{figure}
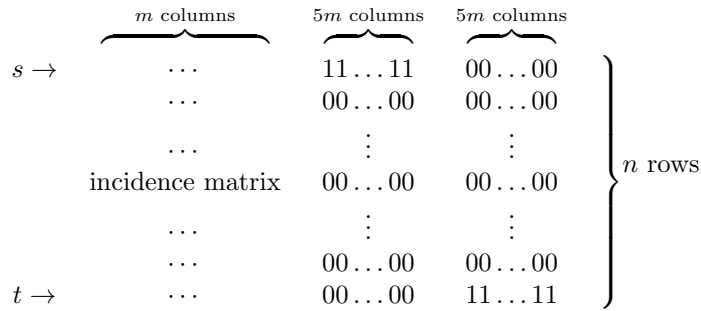
\centering
\begin{minipage}[c]{.65\textwidth}
\begin{tabular}{ccccc}
              &   $\overbrace{~~~~~~~~~~~~~~~~~~~}^{m\textrm{~columns}}$    & $\overbrace{~~~~~~~~~~}^{5m\textrm{~columns}}$  &   $\overbrace{~~~~~~~~~~}^{5m\textrm{~columns}}$ & \\
$s \rightarrow $& $\cdots$ & $1 1 \dots 1 1$    &  $0  0 \dots 0 0$ & \multirow{7}{*}{$\left \}\begin{matrix}~\\~\\~\\ n \textrm{~rows}\\~\\~\\~\\~\\\end{matrix} \right .$}\\
 &  $\cdots$  & $0  0 \dots 0 0$ & $0  0 \dots 0 0$&  \\
 &  $\cdots$  & $\vdots$ & $\vdots$ & \\ 
  & \textrm{incidence matrix} & $0  0 \dots 0 0$ & $0  0 \dots 0 0$ & \\
 &  $\cdots$  & $\vdots$ & $\vdots$ & \\ 
 &  $\cdots$  & $0  0 \dots 0 0$ & $0  0 \dots 0 0$ & \\
$t \rightarrow $& $\cdots$  &  $0  0 \dots 0 0$ &  $1 1 \dots 1 1$ &  \\
\end{tabular}   

 \end{minipage}
\caption{\label{fig:proof1}Matrix described in the proof of Lemma~\ref{lemma:runcountnphard}}
\end{figure}

Suppose that the row corresponding to $s$ is not first or last. Then the number of
runs in the newly added $10m$~columns is at least $3\times 5 m + 2 \times 5m = 25m$ 
(or $30m$ if both $s$ and $t$ are neither first nor last).
Meanwhile, the number of runs in the original incidence matrix is less than $5m$. Thus, any row order
minimizing the number of runs will have the rows corresponding to $s$ and $t$ first and last.
Without loss of generality, we assume $s$ is first.

A minimum-run solution is obtained from a Hamiltonian path from $s$ to $t$ by putting rows
into the order they appear along the path. Such a solution has two columns with two runs, $n-3$ columns
with three runs (or $n-2$  columns with three runs, if $(s,t)$ were an edge of $G$), and the columns for
the other edges 
in $G$ each have five runs.  
Finally, the $10m$ added columns have two runs each. 
Yet   having so few runs 
implies that an $s$--$t$  Hamiltonian path exists. 
Hence, we have reduced the $s$--$t$ Hamiltonian path problem to minimizing the \textsc{RunCount}
by row reordering. 
\end{proof}

\section{Lexicographic and Gray-code sorting}
\label{sec:lex-and-gc-sorting}

While the row reordering problem is NP-hard, 
sorting
is an effective heuristic to enhance column-oriented
indexes~\cite{1083658,arxiv:0901.3751}.
Yet there are many ways to sort rows.

A total order over a set is such that it is transitive ($a\leq b$ and $b \leq c$ implies $a\leq c$), antisymmetric ($a\leq b$ and $b\leq a$ implies $a=b$) and total ($a\leq b$ or $b\leq a$). 
A list of tuples is \emph{discriminating}~\cite{cai1995umd} if all duplicates are
listed consecutively. Orders are discriminating.

We consider sorting functions over tuples. We say that an order over $c$-tuples 
\emph{generates} an order over $c-1$-tuples if and only if the projection of
all sorted lists of $c$-tuples on the first $c-1$ components is discriminating.
When this property applies recursively, we say that we have a recursive order:

\begin{definition}
A \emph{recursive} order over $c$-tuples is such that it generates a recursive order over $c-1$-tuples.
All orders over 1-tuples are recursive.
\end{definition}

An example of 
an order  
 that is \emph{not} recursive is (1,0,0), (0,1,1),
(1,0,1), since its projection on the first two components is not discriminating:  (1,0), (0,1), (1,0).
 We consider several recursive orders
, including
lexicographic order and two Gray-code orders.

\paragraph{Lexicographic order}

\begin{figure*}[tbh]
\centering
\subfloat[Lexicographic\label{fig:lexicopicture}]{\includegraphics[height=0.45\textwidth,angle=270]{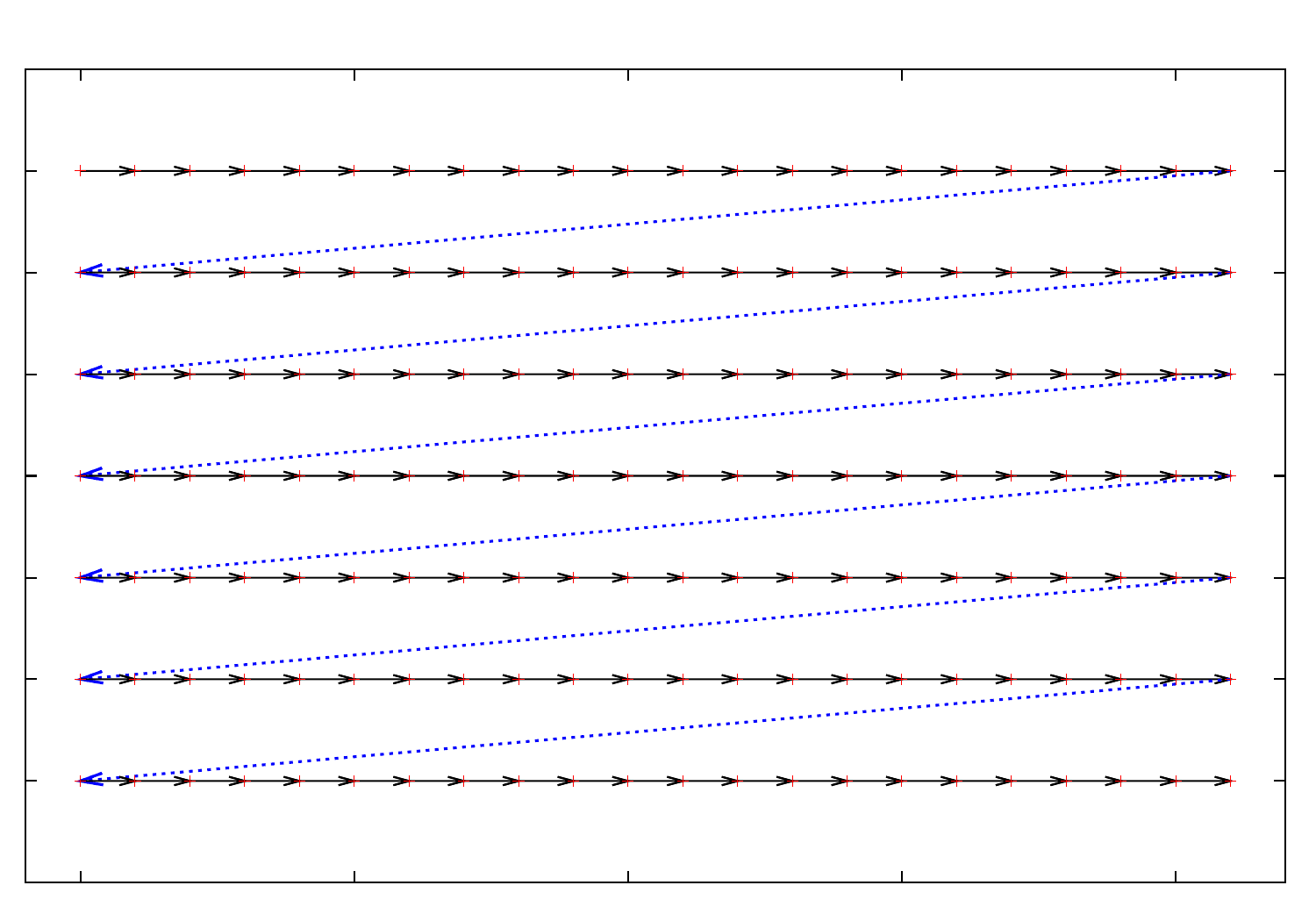}}
\subfloat[Reflected Gray-code\label{fig:reflectedpicture}]{\includegraphics[height=0.45\textwidth,angle=270]{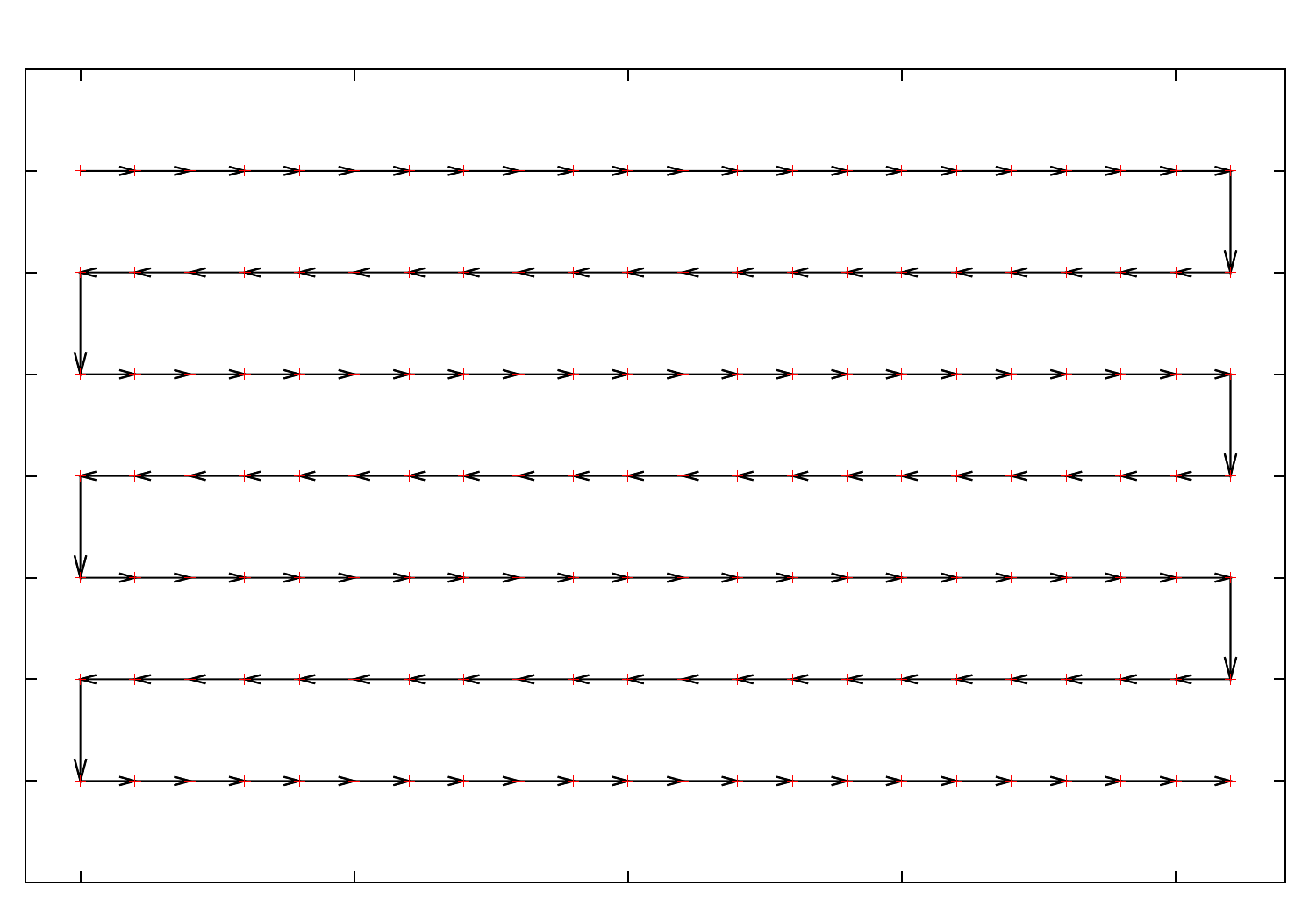}}\\
\subfloat[Modular Gray-code\label{fig:modalpicture}]{\includegraphics[height=0.45\textwidth,angle=270]{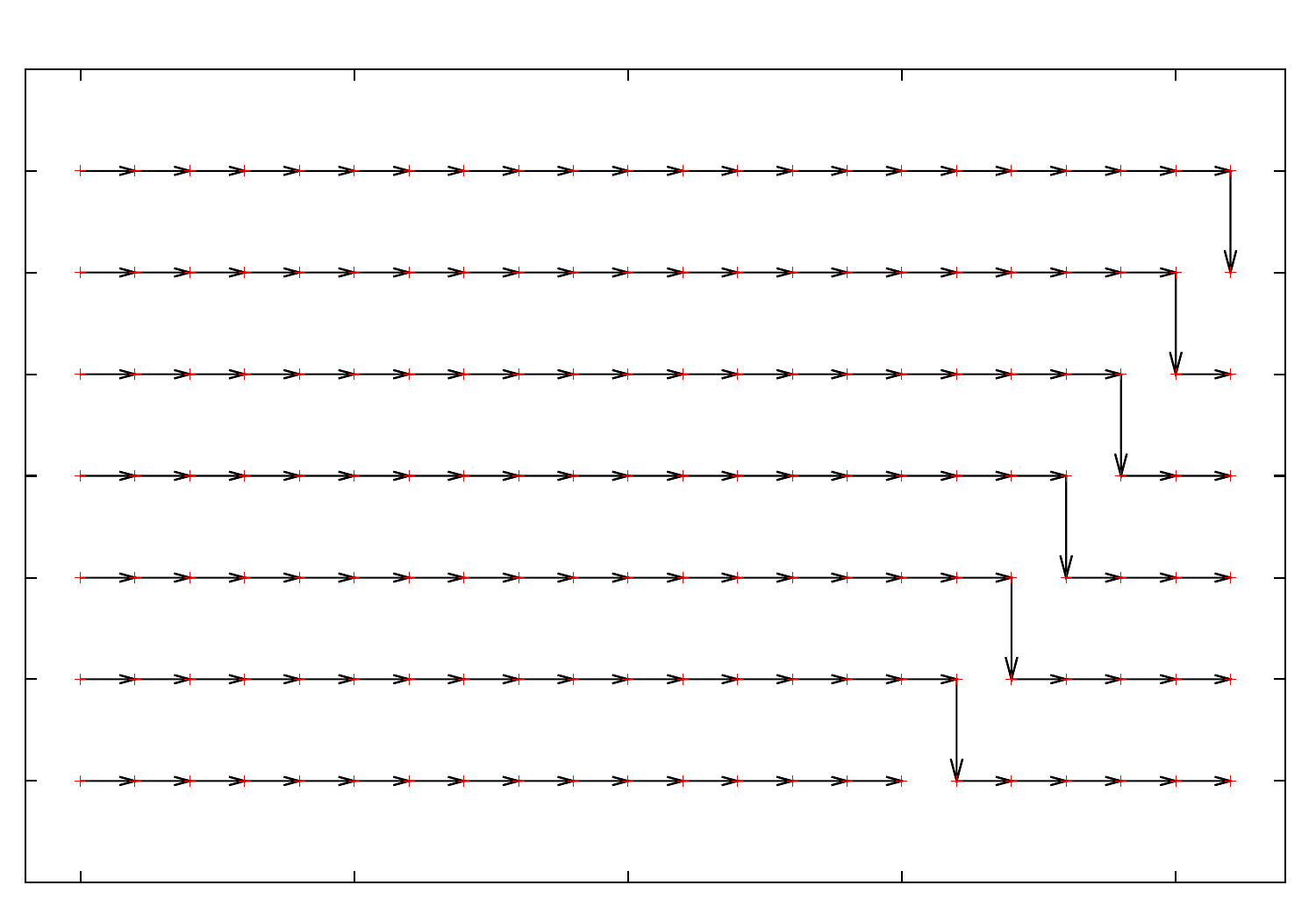}}
\subfloat[Compact Hilbert Index\label{fig:hilbertpicture}]{\includegraphics[height=0.45\textwidth,angle=270]{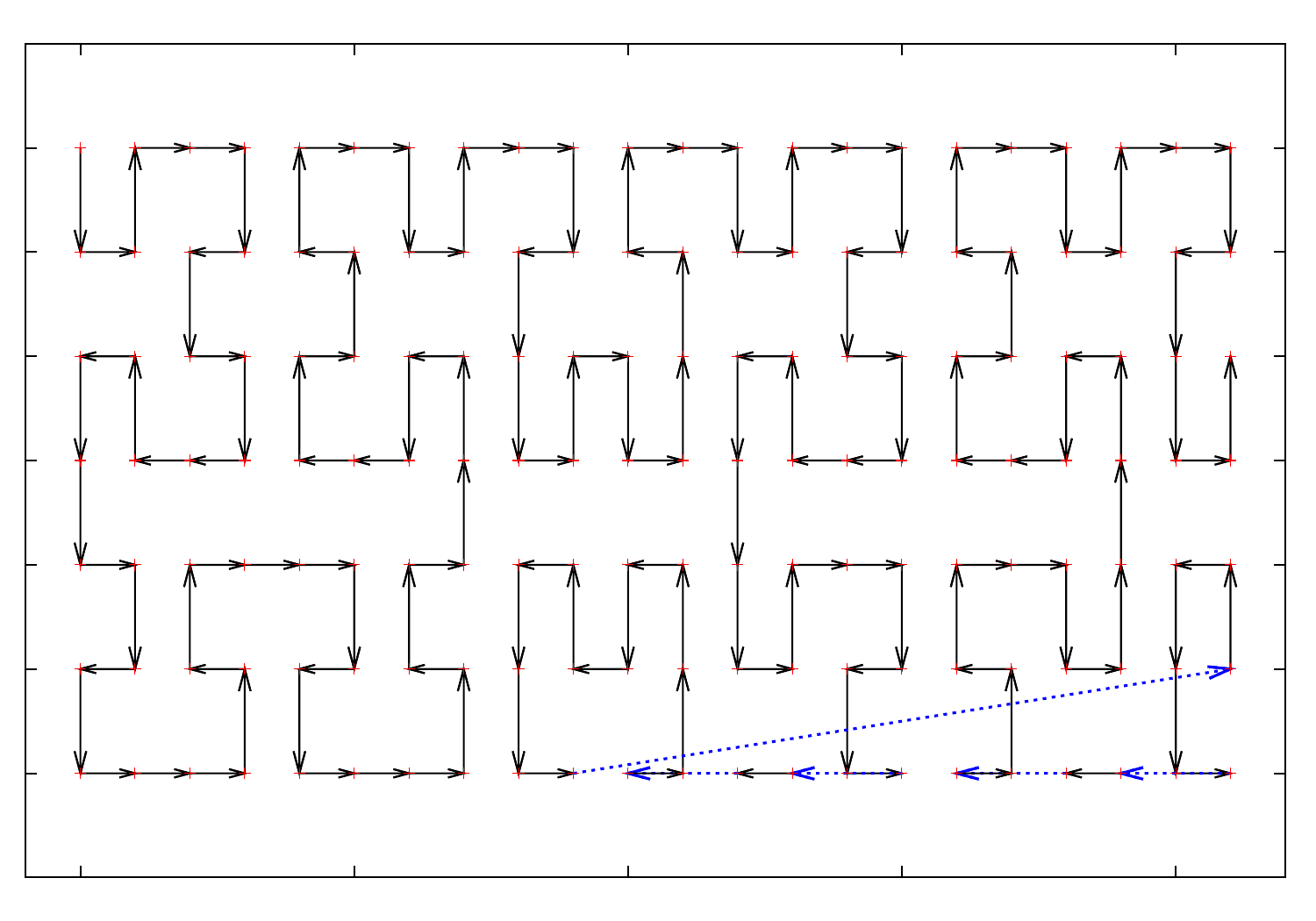}}
\caption{\label{fig:variousorderings}Various orderings of the points in a two-dimensional array}
\end{figure*}

The lexicographic order is also commonly known as the dictionary order.
When comparing two tuples $a$ and $b$, we use the first 
component where they differ ($a_j\neq b_j$ but $a_i=b_i$ for $i<j$)
to decide which tuple is smaller (see Fig.~\ref{fig:lexicopicture}).

Let $N_i$ \label{Ni-defined}be the cardinality of column $i$ and $n$ be the number of rows.
\label{defn-n1c}Given all possible $N_{1,c}\equiv \prod_{i=1}^c N_i$~tuples, we have $N_{1,c}$~runs in the last
column, $N_{1,c-1}$~runs in the second last column and so on. Hence,
we have a total of 
$\sum_{j=1}^c N_{1,j}$~runs. 
If the $N_i$'s have the same value $N_i=N$ for all $i$'s, then we have $N^c+N^{c-1}+\cdots+N = \frac{N^{c+1}- 1}{N-1}-1$~runs. 

\paragraph{Gray-code orders}

We are also interested in the more efficient Gray-code orders. A Gray code is a list
of tuples such that the Hamming distance---alternatively the Lee metric~\cite{1312181}---between successive tuples is one~\cite{Chen2011620,Fang20084679}.
Knuth~\cite[pp.~18--20]{knut:vfour-fascicle-two} describes two types of decimal 
Gray codes. 
\begin{itemize}
\item Reflected Gray decimal ordering is such that
each digit goes from 0 to 9, up and down alternatively:
000, 001, \ldots, 009, 019, 018, \ldots, 017, 018, 028, 029, \ldots, 099, 090, \ldots
\item Modular Gray decimal is such that digits always
increase from 1 modulo 10:
000, 001, \ldots, 009, 019, 010,\ldots, 017, 018, 028, 029, 020, \ldots
\end{itemize}
The extension to the mixed-radix case~\cite{1312181,savage1997scg} from the  decimal codes 
is straight-forward~\cite{richards1986dca} (see Figs.~\ref{fig:reflectedpicture} and~\ref{fig:modalpicture}). 

\begin{figure}
\centering \begin{tabular}{ccc}
 \textbf{0} &  \textbf{0} &  \textbf{0}\\
0 & 0 & \textbf{1}\\
0 &  \textbf{1} & 1\\
0 & 1 &  \textbf{0}\\
 \textbf{1} & 1 & 0\\
1 & 1 &  \textbf{1}\\
1 &  \textbf{0} & 1\\
1 & 0 &  \textbf{0}\\
 \textbf{2} & 0 & 0\\
2 & 0 &  \textbf{1}\\
2 &  \textbf{1} & 1\\
2 & 1 &  \textbf{0}\\
\end{tabular}
\caption{\label{fig:explaingray}A table sorted in  a (reflected) Gray-code order. 
Except for the first row, there is exactly one new run initiated
in each of the $N_{1,c}$~rows (in bold). 
Thus, the table  has
$c-1+N_{1,c} = 3 -1 + 3 \times 2 \times 2 = 14$~column runs. 
}
\end{figure}

Because the Hamming distance between successive codes is one, if 
all possible $N_{1,c}$~tuples are represented, there are
exactly $c-1+N_{1,c}$~runs. If $N_i=N$ for all $i$, then
we have $c-1+ N^c$~runs (see Fig.~\ref{fig:explaingray}). 
All recursive Gray-code orders have $N_1$~runs in the first column,
$N_1 N_2 - N_1 + 1$ runs in the second column, and
the number of runs in column $j$ is given by 
\begin{eqnarray}\label{formula:graycode}
r_j=1+ (N_j-1) N_{1,j-1}.
\end{eqnarray}
(Being a recursive order, the values from the first $j-1$ columns
form $N_{1,j-1}$ blocks, where rows in each block agree on their first
$j-1$ components.  Being a Gray-code order, at any transition
from one block to the next,  values in column $j$ must match.)
If we assume $N_i>1$ for all $i\in \{1,\ldots, c\}$,
then later
columns always have more runs.

From a software-implementation point of view, the lexicographic order
is more convenient than the reflected and modular Gray codes.
A common approach to sorting large files in external memory is divide
the file into smaller files, sort them, and then merge the result.
Yet, with these Gray codes, it is not possible to sort the smaller files
independently: a complete pass through the entire data set may be required
before sorting.
Indeed, consider the these two lists sorted in reflected Gray-code order:
\begin{itemize}
\item Anna  Awkland, Anna  Bibeau, Greg   Bibeau, Greg   Awkland;
\item Bob  Awkland, Bob  Bibeau.
\end{itemize}
Because we sorted the first list without knowing about the first name ``Bob''
a simple merging algorithm  fails. For this
reason, it may be faster to sort data by the lexicographic order.

Similarly, while a binary search through a lexicographically sorted list only
requires comparing individual values (such as Bob and Anna),  binary
searches through a reflected or modular Gray-code ordered list may require the complete
list of values in each column. 

\paragraph{Non-recursive orders}
There are 
balanced and nearly balanced Gray codes~\cite{knut:vfour-fascicle-two,flahive2007bcr,flahive2008bcr}. 
Unlike the other types of Gray codes, the number of runs in all columns
is nearly the same when sorting all possible tuples for $N_1=N_2=\ldots=N_c$. However, they cannot be recursive.
Some authors have used Hilbert space-filling curves to order data
points~\cite{kamel1994hrt,hamilton2007chi,HaverkortW09} (see
Fig.~\ref{fig:hilbertpicture}).  This order is not recursive. Indeed,
the following 2-tuples are sorted in Hilbert order: (1,1),
(2,1),(2,2), (1,2). Yet their projection on the first component is
not discriminating: 1, 2, 2, 1.  It is a balanced Gray code when all
column cardinalities are the same power of two~\cite{hamilton2007chi}.
Beyond two dimensions, there are many possible orders based on Hilbert
curves~\cite{alber2000mch}.  There are also many other alternatives
such as Sierpi\'nski-Knopp order, Peano order~\cite{peano1890courbe},
the Gray-coded curve~\cite{16877}, Z-order~\cite{lebesgue1904lecons}
and H-index~\cite{niedermeier2002tol}. They are often selected for
their locality properties~\cite{1431053}.

If not balanced, non-recursive orders can be \emph{column-oblivious} if the number of runs
per column is independent of the order of the columns. As a trivial example, if you reorder
the columns before sorting the table lexicographically, then the initial order of the columns
is irrelevant.


\subsection{Significance of column order} 
\label{sec:significanceofcolumnorder}
Recursive orders 
depend on the column order. For lexicographic or reflected Gray-code
orders, permuting the columns generates a new row ordering.
The next proposition shows that the effect of the
column ordering grows linearly with the number of columns.

\begin{proposition}\label{prop:neverworsethanc}For tables with $c$~columns, the number of column runs after the application of any recursive-order function can vary by a factor arbitrarily close to $c$ under the permutation of the columns.
\end{proposition}
\begin{proof}
The proof is by construction. 
Given a recursive-order function, we find a $c$-column table that has many
runs when processed by that function.  However, swapping any column with the
first yields a table that---recursively sorted in any way---has few runs.

Consider a column made of $n$~distinct values, given in sorted order: 
A, B, C, D, \ldots
%
 This column is the first column of a $c$-column table. For every odd row, fill all remaining columns with the value 0, and every even row with the value 1:
\begin{eqnarray*}
\begin{tabular}{cccc}
A & 0 & $\cdots$ & 0 \\
B & 1 & $\cdots$ & 1 \\
C & 0 & $\cdots$ & 0 \\
D & 1 & $\cdots$ & 1 \\
$\vdots$ & $\vdots$ & $\cdots$ & $\vdots$ \\
\end{tabular}  
\end{eqnarray*}
This table 
has $n c$ runs and 
is already sorted.
But putting any other column first, any recursive order reduces the number of runs to $n+2(c-1)$. For $n$ large, $\frac{cn}{n+2(c-1)}\rightarrow c$ which proves the result. 
\end{proof}

The 
construction
in the proof uses a high cardinality column. However, we
could replace this single high cardinality column by $\lceil \log_N n \rceil$~columns
having a cardinality of at most $N$, and the result would still hold.


Hence, recursive orders can generate almost $c$~times more
runs than an optimal order.
Yet no row-reordering heuristic can generate more than $c$~times the number of column runs than the optimal ordering solution: there are at least $n$~column runs given $n$~distinct rows, and no more than $c n$~column runs in total. Hence---as row-reordering heuristics---recursive 
orders have 
no useful
worst-case guarantee over arbitrary tables. 
We shall show that the situation differs when column
reordering is permitted.

Suppose we consider a sorting algorithm that first applies a known reordering to columns, then
applies some recursive-order function.  The proposition's bound still applies, because we can 
make an obvious modification to the construction, placing the non-binary column in a 
possibly different position.    The next refinement might be to consider a sorting algorithm
that---for a given table---tries out several different column orders.  For the construction
we have used in the proof, it  always finds an optimal ordering.

Unfortunately, even allowing the enumeration of all possible column reorderings is insufficient to make recursive 
ordering optimal.
 Indeed, consider the 
table in Fig.~\ref{fig:reordering-recursive-insufficent}.
The Hamming distance between any two consecutive tuples is one. Thus each new row initiates exactly one new run, except for the first row.  
Yet, because all tuples are distinct, this is a minimum: a Hamming distance of zero is impossible.
Thus, this row ordering has a minimal number of column runs.
We prove that no recursive ordering can be similarly optimal.

%
%
%

\begin{figure}\centering
\begin{tabular}{|cc|}
 K & Y\\
 A & Y\\
 A & D\\
 Z & D\\
 Z & B\\
 A & B\\
 A & C\\
 W & C\\
 W & E\\
 F & E\\
 F & C\\
 H & C \\
 H & J\\
\end{tabular}  
\caption{A table such that no recursive ordering is optimal.}
\label{fig:reordering-recursive-insufficent}
\end{figure}
%
%

We begin by analyzing the neighbors of a tuple, where
two tuples are neighbors if they have a Hamming distance of one:
\begin{itemize}
\item The tuple (K,Y) has only one neighbor: (A,Y).
\item The tuple (H,J) has only one neighbor: (H,C).
\item The tuples of the form (Z,$\cdot$) only have neighbors of the form (A,$\cdot$).
\item The tuples of the form ($\cdot$, E) only have neighbors of the form ($\cdot$,C).
\end{itemize}
In effect, we must consider all Hamiltonian paths in the graph of neighbors.
The ordered list must begin and end with (K,Y) and (H,J), if it is optimal.
A recursive order must be discriminating on the first column.
Without loss of generality, suppose that the list begins by (K,Y).
Thus, all tuples of the form (A,$\cdot$) must follow by recursivity.
Then tuples of the form (Z,$\cdot$) must follow. At this point, we cannot
continue the list by jumping from neighbor to neighbor.
Hence, no recursive ordering is optimal.
A similar argument shows that flipping the two columns leads to the
same result: no recursive ordering can be optimal.

\begin{lemma}
There are tables where no recursive order minimizes the number of runs---even after reordering the columns.
\end{lemma}

Determining a tight bound on the
suboptimality of recursive ordering remains open. 
Recursive orders applied to the example of Fig.~\ref{fig:reordering-recursive-insufficent}
generate at least 15~runs whereas 14~runs is possible, for a suboptimality ratio of $\frac{15}{14}$.
If we allow arbitrarily long two-dimensional tables, we can generalize our construction
to obtain ratios arbitrarily close to $\frac{13}{12}$.  %
Thus, the suboptimality ratio of recursive orders  ranges between 
$\frac{13}{12}$  and $c$.
However, a computer search  through 100,000~uniformly distributed 
tridimensional 
tables with 10~rows and six distinct column values 
failed to produce a single case where recursive ordering is suboptimal.
That is, among the row orderings minimizing the number of runs, at least
one is recursive after some reordering of the columns. 
Hence, it is possible that recursive ordering is rarely suboptimal.

The next proposition gives a simple suboptimality bound on any recursive order.
This result implies that recursive ordering is 3-optimal or better for several 
realistic tables (see Table~\ref{tab:caractDataSet}). 

\begin{proposition}\label{prop:optimalitybound}\label{mu-defined}
Consider a table with $n$~distinct rows and column cardinalities $N_i$ for $i=1,\ldots,c$.
Recursive ordering is $\mu$-optimal for the problem of minimizing the
runs where \begin{eqnarray*}
\mu & = &  \frac{\sum_{j=1}^c \min(n,N_{1,j}) }{n+c-1}.
\end{eqnarray*}
The bound $\mu$ can be made stronger if the recursive order is a Gray code:
\begin{eqnarray*}
\mu_{\textrm{GC}} & = &  \frac{\sum_{j=1}^c\min( n, 1+(N_j-1)N_{1,j-1})}{n+c-1}.
\end{eqnarray*}
but $\mu_{\mathrm{GC}} > \frac{\min_i(N_i-1)}{\min_i(N_i)} \mu$
\end{proposition}

\begin{proof}
Given a table in any recursive order, the number of runs in the $i^{\mathrm{th}}$~column
is bounded by $N_{1,i} $ and by $n$. Thus the number of runs
in the table is no more than  $\min(N_1, n) +\min(N_{1,2} , n) + \cdots + \min(N_{1,c} , n)$.
Yet  there are at least $n+c-1$~runs in the optimally-ordered table. Hence, the result follows.
The tighter bound for Gray-code orders follows similarly, and the relationship between
$\mu_{\mathrm{GC}}$ and $\mu$ follows by straightforward algebra ($\mu_{\mathrm{GC}} >\sum_{j=1}^c \min( \beta n, \beta N_{1,j})$
where $\beta = \frac{ \min_i(N_i)-1}{\min_i(N_i)}$).  
\end{proof}

As an example, consider the list of all dates (month, day, year) for a century ($N_1=12, N_2=31, N_3=100, n=12 \times 31 \times 100$): then $\mu \approx 1.01$
so that lexicographic sorting is within 1\% of minimizing the number of runs.
The optimality bound given by Proposition~\ref{prop:optimalitybound} is tighter when the columns
are ordered in non-decreasing cardinality ($N_1\leq N_2 \leq \cdots \leq N_c$). This fact alone
can be an argument for ordering the columns in increasing cardinality.

\subsection{Determining the optimal column order is NP-hard}
\label{sec:colr}

For lexicographic sorting, it is NP-hard to determine which column ordering
will result in least cost under the \textsc{RunCount} model, even when the tables have only two values.
We consider the following decision problem:

\paragraph{Column-Ordering-for-Lex-Runcount (COLR)}
 Given table $T$ with binary values 
and given  integer $K$, is there a column ordering such that the
lexicographically sorted $T$ has at most $K$~runs? 


\begin{theorem}
COLR is NP-complete.
\end{theorem}
\begin{proof}
Clearly the problem is in NP\@.  Its NP-hardness is shown by reduction from the
variant of Hamiltonian Path where the starting vertex is given~\cite[GT39]{gare:gandj}.
Given an instance $(V,E)$ of Hamiltonian Path, without loss of
generality let $v_1 \in V$ be the specified starting vertex.   
We construct a table $T$ as follows:
first, start with the incidence matrix.  Let $V = \{v_1, v_2, \ldots , v_{|V|}\}$
and $E = \{ e_1, e_2, \ldots, \ldots, e_m\}$. Recall that this matrix has a column for each
edge and a row for each vertex; $a_{i,j}=1$ if edge $e_j$ has vertex $v_i$ as an endpoint and otherwise $a_{i,j}=0$.  Vertex $v_1$ corresponds to the first row. 
We prepend and append a row of zeros to the incidence matrix.
Next we prepend $h$ columns with values $10^{|V|+1}$ (i.e., $100\dots0$) and
$h$ columns with $110^{|V|}$; 
see Fig.~\ref{fig:ham-path-reduction} for an example.  The value of $h$ is ``large'';
we compute the exact value later.

\newcommand{\oneb}{\cellcolor{blueish}{1}}
\newcommand{\zerob}{\cellcolor{blueish}{0}}

\begin{figure}
\begin{centering}
\subfloat[Graph]{\begin{minipage}[c]{.3\textwidth}
\vspace{1.6cm} 
\includegraphics[width=1\textwidth]{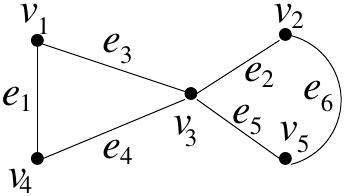}
\end{minipage}
}
\qquad
\subfloat[Constructed table, incidence matrix shaded.]
{
\begin{minipage}[c]{.65\textwidth}
\begin{tabular}{ccccccccc}
              &               & $e_5$ & $e_1$  & $e_2$   & $e_3$  & $e_4$   &  $e_6$ &                \\
$ 1 \cdots 1$ & $1 \cdots 1$  & 0     &  0     &  0      &  0     &  0     &   0    &                \\
$ 0 \cdots 0$ & $1 \cdots 1$  &\zerob & \oneb  & \zerob  & \oneb  & \zerob &  \zerob& {\small $v_1$} \\
$ 0 \cdots 0$ & $0 \cdots 0$  &\zerob & \zerob & \oneb   & \zerob & \zerob &  \oneb & {\small $v_2$} \\
$ 0 \cdots 0$ & $0 \cdots 0$  &\oneb  & \zerob & \oneb   & \oneb  & \oneb  &  \zerob& {\small $v_3$} \\
$ 0 \cdots 0$ & $0 \cdots 0$  &\zerob & \oneb  & \zerob  & \zerob & \oneb  &  \zerob& {\small $v_4$} \\
$ 0 \cdots 0$ & $0 \cdots 0$  &\oneb  & \zerob & \zerob  & \zerob & \zerob &  \oneb & {\small $v_5$} \\
 $\underbrace{0 \cdots 0}_h$
              & $\underbrace{0 \cdots 0}_h$  
                              & 0     &  0     &  0      &  0     &  0     &  0     &                \\ 
 \end{tabular}   
 \end{minipage}
}
\end{centering}
\caption{\label{fig:ham-path-reduction} Table built from graph on the left. 
There
are $h$ copies of the column that begins $10\ldots$ and the column that begins with $11\ldots$.
}
\end{figure}

We show the resulting instance, with table $T$ and bound $K = 4h+3(|V|-1)+5(m-|V|+1))$,
satisfies the requirements for 
COLR if and only if
$(V,E)$ contains a Hamiltonian path starting at $v_1$. 

First, suppose that we have a suitable Hamiltonian path
in $(V,E)$. Let $\epsilon_i \in E$ be the $i^{\mathrm{th}}$ edge along this path.
Edge $\epsilon_1$ is incident upon $v_1$.

Reorder the columns of $T$: leave the first $2h$~columns in their current order.
Next, place the columns corresponding to $\epsilon_i$
in order $\epsilon_1 , \epsilon_2, \ldots , \epsilon_{|V|-1}$. 
The remaining columns follow in an arbitrary order.
See Fig.~\ref{fig:constructed-table-lex-sorted}, where it is apparent that
the constructed table is already lexicographically sorted\footnote
{We sort with 1 ordered before 0.}
Also, 
the first $2h$ columns have 2~runs each, and  the $|V|-1$~columns for
$\epsilon_i$ have three runs each (each has the value $0^i110^{|V|-i}$). 
 The remaining $m-|V|+1$~columns have five
runs each%
:
all patterns with adjacent ones have been used (and 
there are no duplicates); hence, all remaining patterns are of the form
$0^+10^+10^+$.  Thus the bound is met.

\renewcommand{\zerob}{0}
\renewcommand{\oneb}{1}
\begin{figure}
\begin{centering}
{\small 
\begin{tabular}{ccccccccc}
              &              & $e_1$  & $e_4$ & $e_2$   & $e_6$  &$e_3$   & $e_5$   &                \\
 $1 \cdots 1$ & $1 \cdots 1$ &  0     &  0    &  0      &  0    &  0      &  0     &                \\
 $0 \cdots 0$ & $1 \cdots 1$ & \oneb  & \zerob& \zerob  & \zerob& \oneb   & \zerob & {\small $v_1$} \\
 $0 \cdots 0$ & $0 \cdots 0$ & \oneb  & \oneb & \zerob  & \zerob& \zerob  & \zerob & {\small $v_4$} \\
 $0 \cdots 0$ & $0 \cdots 0$ & \zerob & \oneb & \oneb   & \zerob& \oneb   & \oneb  & {\small $v_3$} \\
 $0 \cdots 0$ & $0 \cdots 0$ & \zerob & \zerob& \oneb   & \oneb & \zerob  & \zerob & {\small $v_2$} \\
 $0 \cdots 0$ & $0 \cdots 0$ & \zerob & \zerob& \zerob  & \oneb & \zerob  & \oneb  & {\small $v_5$} \\
 $0 \cdots 0$ & $0 \cdots 0$ &  0     &  0    &  0      & 0     &  0      &  0     &                \\ 
 \end{tabular}\\                                              
}
\end{centering}
\caption{\label{fig:constructed-table-lex-sorted} A lexicographically sorted table with the
required \textsc{RunCount} bound is obtained from the Hamiltonian path consisting
of edges $e_1, e_4, e_2, e_6$.}
\end{figure}

Next, suppose $T$ satisfies the requirements of COLR with the given bound 
$K=4h+3(|V|-1)+5(m-|V|+1)$.
We show this implies $(V,E)$ has a Hamiltonian path starting with $v_1$.

If $h$ is large enough, we can guarantee that the first two rows 
have not changed their initial order.
This is enforced by the $2h$ columns that were initially
placed leftmost.  Their column values must end with 0 (the row
of zeros is always last after lexicographic sorting
).  If we
analyze the \textsc{RunCount} cost of these columns, they cost
$4h$ when the first two rows remain in their initial order, otherwise
they cost $5h$ or $6h$. If $h$ is large enough, this penalty
will outweigh any possible gain from having a column order that,
when sorted, moves the first two rows. 

Knowing the first row, we deduce that every column begins
with a one if it is one of the $2h$~columns, but it begins with a zero in every
remaining column.
We now focus on these remaining columns, which correspond to edges in $E$.
Since each column value
begins and ends with zero and has exactly two
ones,  its pattern is either $0^+110^+$ (3 runs)  or $0^+10^+10^+$ (5 runs).
The specified \textsc{RunCount} bound
implies that we must have $|V|-1$ columns with 3 runs.  The edges for
these columns form the desired Hamiltonian path that starts at $v_1$.

To finish, we must choose $h$ such that the penalty
(for choosing a column ordering that disrupts the order of the first two
rows after lexicographic sorting) exceeds any possible gain.  The increased
cost from $4h$ is at least $5h$, a penalty of at least $h$.
An upper bound on the gain from the other columns is $3m$ because the
\textsc{RunCount} is no more than $5m$ and cannot be decreased
below $2m$. Choose $h = 3m+1$.
\end{proof}

This result can be extended to the reflected Gray-code order and, we conjecture,
to all recursive orders. 
A related problem tries to minimize the maximum number of runs in any table 
column.
This problem is also NP-hard (see Appendix~\ref{sec:COMLR}).

Moreover, given a very large number of rows, it might impractical to
try more than one column order. Indeed, evaluating
each new solution implies sorting the table, a potentially expensive step.
Thus, heuristics which only consider a few easily computed statistics, such as
cardinality, are preferable.

\section{Increasing-cardinality-order minimizes runs}
\label{sec:Increasing-cardinality-order}


Consider a sorted table. The table might be sorted in 
lexicographic order or in reflected Gray-code order.
Can we prove that sorting the columns in increasing cardinality is a sensible
heuristic to minimize the number of runs?
We consider analytically two cases: (1)~complete tables 
and (2)~uniformly distributed tables.


\subsection{Complete tables}
\label{sec:complete-tables}


Consider a $c$-column table with column cardinalities
$N_1, N_2,\ldots, N_c$. 
A \emph{complete} table is one where all $N_{1,c}$ possible tuples
are present.
In practice, even if a table is not complete, the projection 
on the first few columns might be complete.
Using a lexicographic order,
a complete table has $\sum_{j=1}^c N_{1,j}$~runs, hence
the \textsc{RunCount} is minimized when the columns
are ordered in non-decreasing 
  cardinality: $N_i \leq N_{i+1}$ for 
$i=1,\ldots, c-1$. 
Using Gray-code ordering, a complete table has only 
 $c-1+N_{1,c}$~runs (the minimum possible) no
matter how  the columns are ordered. Hence,
for Gray-code, the  \textsc{RunCount} of complete tables is
not sensitive to the column order.

Somewhat artificially, we can create a family of recursive orders for which 
the \textsc{RunCount} is not minimized over complete tables when the columns
are ordered in increasing cardinality. Consider
the following family: ``when $N_1$ is odd, use reflected Gray 
code order.  Otherwise, use lexicographic order.'' For $N_1=2$
and $N_2=3$, we have 8~runs using lexicographic order.
With $N_1=3$ and $N_2=2$, we have 7~runs using any recursive
Gray-code order. 
Hence, we cannot extend our analysis to all families of 
recursive orders from Gray-code and lexicographic orders.
Nevertheless, if we assume that all column cardinalities are large, then
the number of runs tends to $N_{1,c}$ and all column orders become equivalent.

\begin{figure}
\centering
\includegraphics[width=0.7\columnwidth]{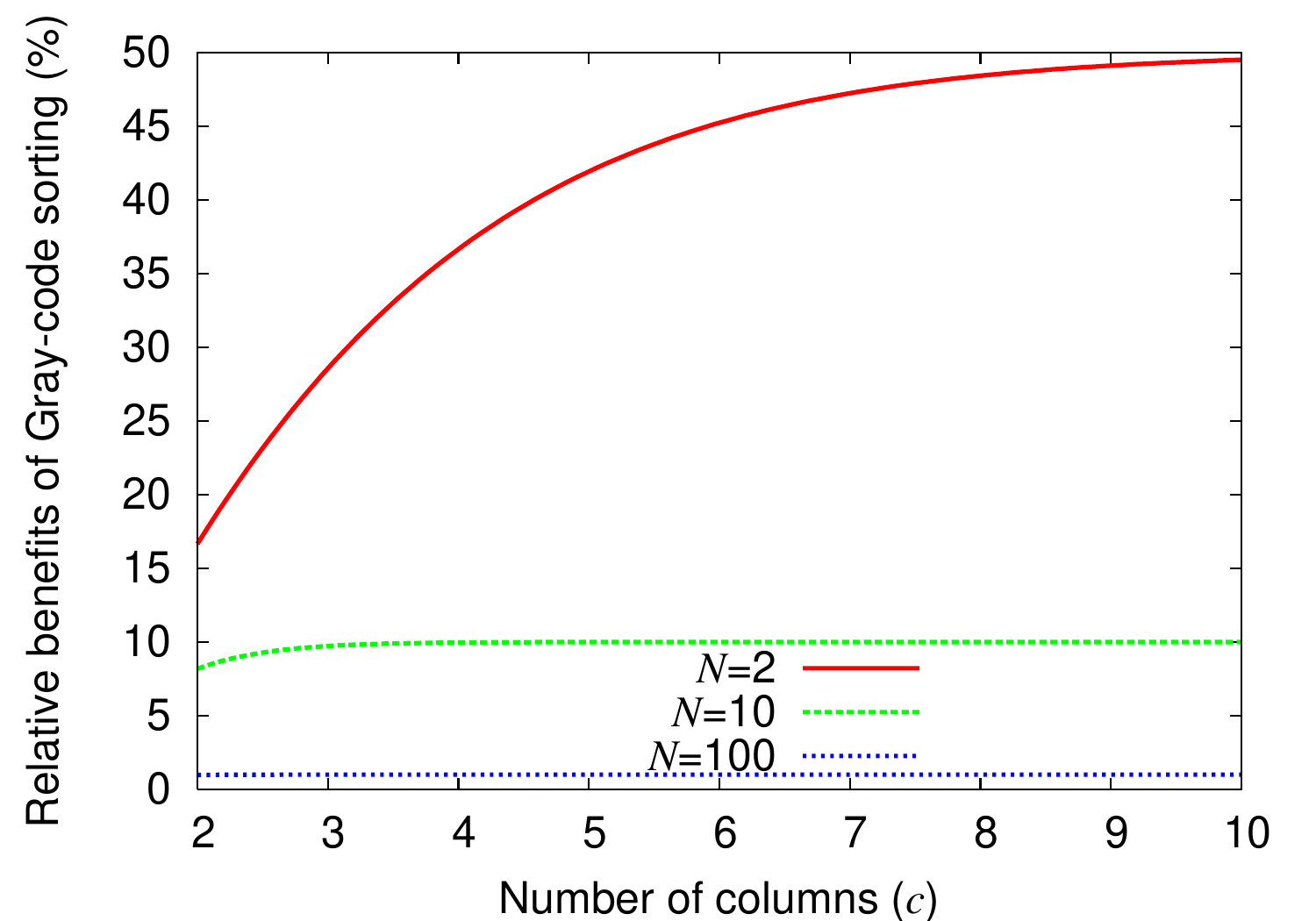}
\caption{\label{fig:relativebenefitsgray}Relative benefits of Gray-code sorting against lexicographic orders for complete $c$-column table where all
column cardinalities are $N$}
\end{figure}

The benefits of Gray-code orders---all Gray-code orders, not just recursive Gray-code
orders---over lexicographic orders are small for complete tables having high cardinalities as the next proposition shows (see Fig.~\ref{fig:relativebenefitsgray}).

\begin{proposition}Consider the number of runs in complete tables with columns having cardinality $N$. The relative
benefit of Gray-code orders over lexicographic orders grows monotonically with $c$
and is at most $1/N$.
\end{proposition}
\begin{proof}
The relative benefits of Gray-code sorting for
complete tables with all columns having cardinality $N$ is
$\frac{\frac{N^{c+1}-1}{N-1}- 1 - (N^{c}+c-1)}{\frac{N^{c+1}-1}{N-1}- 1}$.
As $c$ grows, this quantity converges to $1/N$ from below.
\end{proof}

\subsection{Uniformly distributed case}
\label{sec:uniformlydistributed}

We consider tables with column cardinalities $N_1, N_2,\ldots N_c$.
Each of the $N_{1,c}$~possible tuples is present with
probability $p$. When $p=1$, we have complete tables.

For 
recursive orders over uniformly distributed tables, 
knowing how to compute the expected
 number of runs in the second column 
 of a two-column table is almost sufficient to analyze the general case.
  Indeed, given a 3-column
 table, the second column behaves just like the second column in
 2-column table with $p\leftarrow 1-(1-p)^{N_3}$. Similarly,
 the third column behaves just like the second column in a
 2-column table with  $N_1 \leftarrow N_1 N_2$ and $N_2\leftarrow N_3$.
 
This second column is divided into $N_1$~blocks of $N_2$ tuples, each tuple
having a probability $p$ of being present. The expected number
of tuples present in the table is $N_1 N_2 p$. However,
$N_1 N_2 p$
is an overestimate
of the number of runs in the second column%
. We need to subtract the expected number of 
\emph{seamless joins}
 between blocks: two blocks have a seamless join if the first one terminates with the first value of the second block.
The expected number of seamless joins
is no larger than the expected number of non-empty blocks minus one: 
\label{rho-defined}
$N_1 \rho_{N_2} -1$ where $\rho_{N_2} \equiv 1-(1-p)^{N_2}$.
While for complete tables, all recursive Gray-code orders 
agree on the
number of runs and seamless joins per column, the same is not true for 
uniformly distributed tables.
Different recursive Gray-code orders have different expected numbers
of seamless joins.

Nevertheless, we wish to prove a generic result for tables having
large column cardinalities ($N_i \gg 1$ for all $i$'s). 
Consider a two-column table having uniform column cardinality $N$.
For any recursive order, the expected number of seamless joins is less than $N\rho_N$. 
However, the expected
sum of the number of runs and seamless joins is $N \rho_N$ in the first column and $N^2 p$ in the second,
for a total of $N \rho_N + N^2 p$. For a
fixed table density, the ratio $N\rho_N/(N \rho_N + N^2 p)$ goes to zero as $1/N$
since $\rho_N\rightarrow 1$ exponentially.
Hence, for tables having large column cardinalities,
the expected number of seamless joins is negligible compared to the expected
number of runs. The following lemma makes this result precise.

\begin{lemma}\label{lemma:asymptruns}
Let $S_{i}$ and $R_{i}$ be the expected number of seamless joins and runs in column $i$.
For all recursive orders, we have 
\begin{eqnarray*}
\frac{\sum_{i=1}^c S_i}{\sum_{i=1}^c S_i + \sum_{i=1}^c R_i} \leq \frac{1}{\min_{k\in\{1,2,\ldots,c\}}N_k}
\end{eqnarray*}
over uniformly distributed tables.
\end{lemma}
\begin{proof}
Column $i+1$ has an expected total of runs and seamless joins of
$S_{i+1}+R_{i+1}=N_{1,i+1} \rho_{N_{i+2}\ldots N_c}$.
It has less than  $N_{1,i} \rho_{N_{i+1}\ldots N_c}$~seamless joins.
We can verify that $\rho_{N_{i+1}\ldots N_c}\leq \rho_{N_{i+2}\ldots N_c}$ for all $p\in [0,1]$.
Thus $S_{i+1}/(S_{i+1}+R_{i+1}) \leq 1/N_{i+1}$. 

Hence, we have
$  N_i S_i \leq S_i + R_i$. This implies that $  \min_{k\in\{1,2,\ldots,c\}}N_k S_i \leq S_i + R_i$.
Therefore, we have  $  \min_{k\in\{1,2,\ldots,c\}}N_k \sum_{i=1}^c S_i \leq \sum_{i=1}^c S_i + \sum_{i=1}^c R_i$
which proves the result.
\end{proof}


Therefore, for large column cardinalities, we can either consider
the number of runs, or the sum of the runs and seamless joins. 
In this context, the next proposition shows that it is best to order columns in increasing
cardinality. 

\begin{proposition}\label{prop:danielbeingclever}The expected sum  of  runs and seamless joins is the same for all
recursive orders. Moreover,
it is minimized over uniformly distributed tables
if the columns are sorted by increasing cardinality.
\end{proposition}
\begin{proof}
For all recursive orders, the expected number of runs and seamless joins for columns $i$ and $i+1$ is
$N_{1,i} \rho_{N_{i+1}\cdots N_c} +N_{1,i+1}  \rho_{N_{i+2}\cdots N_c}$.
The second term---corresponding to column $i+1$---is invariant 
under a permutation of columns $i$ and $i+1$.
We focus our attention on the first term: $N_{1,i} \rho_{N_{i+1}\cdots N_c}$.
After permuting $i$ and $i+1$, it becomes 
$N_{1,i-1} N_{i+1} \rho_{N_{i}N_{i+2}\cdots N_c}$.

To simplify the notation, rewrite
$\rho_{N_{i+1}\cdots N_c}$ and $\rho_{N_{i}N_{i+2}\cdots N_c}$
as 
$\rho_{N_{i+1}}$ and $\rho_{N_{i}}$
by substituting $\rho_{N_{i+2}\cdots N_c}$ for $p$ and let $i=1$.
Thus, we compare $N_1 \rho_{N_2}$ and $N_2 \rho_{N_1}$.

To prove the result, it is enough to show that
$N_1 \rho_{N_2} < N_2 \rho_{N_1}$ implies
$N_1 < N_2$ for $p\in (0,1]$. 
Suppose that it is not the case: it is possible to have $N_1 \rho_{N_2} < N_2 \rho_{N_1}$
and $N_1>N_2$.
Pick such $N_1, N_2$.
Let $x=1-p$, 
then $N_1 \rho_{N_2} - N_2 \rho_{N_1}$
is $N_1 (1-x^{N_2}) - N_2 (1-x^{N_1})$.
The polynomial is positive for $x=0$ since $N_1>N_2$.
Because $N_1 \rho_{N_2} < N_2 \rho_{N_1}$ is possible (for some value of $x$), the polynomial
must
be negative at some point in $(0,1)$, hence it must
have a root in $(0,1)$.
However, the polynomial has only 3~terms so that
it cannot have more than 2~positive roots (e.g., by Descartes' rule of signs).
Yet it has a root of multiplicity two at $x=1$: 
after dividing by $x-1$, we get
$N_1 (1+x+\cdots+x^{N_2-1}) - N_2 (1+x+\cdots+x^{N_1-1})$ which is
again zero at $x=1$. Thus, it has no such root
and, by contradiction, 
$N_1 \rho_{N_2} \leq N_2 \rho_{N_1}$ implies
$N_1 \leq N_2$ for $p\in (0,1]$. The proof is concluded.
\end{proof}

\begin{theorem}\label{theorem:genericresult}
Given 
\begin{enumerate}
\item the expected number of runs $R^{\uparrow}$ in a table sorted using any recursive order with an ordering of the column
in increasing cardinality
and 
\item $R^{\mathrm{optimal}},$ the smallest possible expected number of runs 
out of all possible recursive orders on the table (with the columns ordered in any way), 
\end{enumerate} 
then 
\begin{eqnarray*}
\frac{R^{\uparrow}-R^{\mathrm{optimal}}}{R^{\uparrow}} \leq \frac{1}{\min_{k\in\{1,2,\ldots,c\}}N_k}
\end{eqnarray*}
over uniformly distributed tables.
That is, for large column cardinalities---$\min_{k\in\{1,2,\ldots,c\}}N_k$ is large---
sorting a table recursively with the columns ordered in increasing cardinality 
is asymptotically optimal.
\end{theorem}
\begin{proof}
Whenever $a\geq b$, then $1-a\leq 1-b$. Applying this idea to the statement of Lemma~\ref{lemma:asymptruns}, we 
have
\begin{eqnarray*}
1-\frac{\sum_{i=1}^c S_i}{\sum_{i=1}^c S_i + \sum_{i=1}^c R_i} \geq 1- \frac{1}{\min_{k\in\{1,2,\ldots,c\}}N_k}
\end{eqnarray*}
or
\begin{eqnarray*}
\sum_{i=1}^c R_i \geq  \frac{\min_{k\in\{1,2,\ldots,c\}}N_k-1}{\min_{k\in\{1,2,\ldots,c\}}N_k}\left (\sum_{i=1}^c S_i + \sum_{i=1}^c R_i \right ).
\end{eqnarray*}
Let $S^{\uparrow}$~and $S^{\mathrm{optimal}}$ be the expected number of seamless joins
corresponding to $R^{\uparrow}$ and $R^{\mathrm{optimal}}$. We
have 
\begin{eqnarray*}
R^{\mathrm{optimal}} & \geq & \frac{\min_{k\in\{1,2,\ldots,c\}}N_k-1}{\min_{k\in\{1,2,\ldots,c\}}N_k} (R^{\mathrm{optimal}} + S^{\mathrm{optimal}})\\
 & \geq  &\frac{\min_{k\in\{1,2,\ldots,c\}}N_k-1}{\min_{k\in\{1,2,\ldots,c\}}N_k} (R^{\uparrow} + S^{\uparrow})~\textrm{\ by Prop.~\ref{prop:danielbeingclever}}\\
 & \geq  &\frac{\min_{k\in\{1,2,\ldots,c\}}N_k-1}{\min_{k\in\{1,2,\ldots,c\}}N_k} R^{\uparrow} 
\end{eqnarray*}
from which the result follows. 
\end{proof}


From this theorem, we can conclude that---over uniformly distributed tables having large
column cardinalities---sorting lexicographically with the column ordered in increasing cardinality
is as good as any other recursive sorting.

The expected benefits of seamless joins are small, at least for uniformly distributed tables.
Yet they cause runs from different columns to partially overlap. Such partial overlaps might
prevent some computational optimizations.   
For this reason, Bruno~\cite{brunoelephant} avoids seamless joins in RLE-compressed columns: each seamless join
becomes the start of a new run. 
In this model, Proposition~\ref{prop:danielbeingclever}
already shows that ordering the columns in increasing cardinality minimizes the
expected number of runs---for uniformly distributed tables. 

	

\subsubsection{Best column order for lexicographic sorting}
\label{sec:lowcardlexico}

While Theorem~\ref{theorem:genericresult} states that the best column ordering---for all 
recursive orders---is by increasing cardinality, the result is only valid asymptotically (for large
column cardinalities).
For the lexicographic order, we want to prove that the best column ordering
is by increasing cardinality, irrespective of the column cardinalities.

The $N_1$~blocks in the second column of a lexicographically ordered are ordered from 1 to $N_2$.
\label{pdd-defined}
Let $P_{\downdownarrows N_2}$ be the probability that any
two non-empty such blocks have a seamless join. 
The probability that the first $x$ tuples in a block are not present
whereas the $x+1^{\mathrm{th}}$~tuple is present is $(1-p)^x p /(1-(1-p)^{N_2})$.
To obtain a seamless join, we need a run
of precisely $N_2-1$ missing tuples, and it can begin anywhere
between the second possible tuple in the first block and the first possible
tuple in the second block.  (See Fig.~\ref{fig:lex-seamless}.)
Hence, we
have $P_{\downdownarrows N_2} = \frac{N_2 p^2 (1-p)^{N_2-1}}{(1-(1-p)^{N_2})^2}
=\frac{N_2 p^2 (1-\rho_{N_2})}{(1-p) \rho_{N_2}^2}$. 
Let $ P_{\downdownarrows N_2,p'}$ and $\rho_{N_2,p'}$ be
$ P_{\downdownarrows N_2}$ and $\rho_{N_2}$ with $p'$ substituted in place of $p$.

\begin{figure}
\hspace*{3cm}\subfloat[Lexicographic\label{fig:lex-seamless}]{ 
$[ 1~2 \cdots k \underbrace{\cdots N_2}_{N_2-k} ][ \underbrace{1~2 \cdots}_{k-1} k \cdots N_2]$
}\\
\hspace*{3cm}\subfloat[Reflected GC (successive blocks)\label{fig:rgc-seamless}]{ 
$[ 1~2 \cdots k \underbrace{\cdots N_2}_{N_2-k} ][ \underbrace{N_2~N_2-1 \cdots}_{N_2-k} k \cdots 1]$
}\\
\hspace*{3cm}\subfloat[Modular GC (successive blocks)\label{fig:mgc-seamless}]{ 
$[ 1~2 \cdots k \underbrace{\cdots N_2}_{N_2-k} ][ \underbrace{N_2~1~2 \cdots}_{k \bmod{N_2}} k \cdots N_2-1]$
}\\
\hspace*{3cm}\subfloat[Modular GC (separated by $y-1$~blocks)\label{fig:mgc-seamless2}]{ 
$[ 1~2 \cdots k \underbrace{\cdots N_2}_{N_2-k} ][\underbrace{s~s+1\cdots}_{(k-1+y) \bmod{N_2}} k \cdots s-1]$
}
\caption{Two consecutive non-empty blocks and the number of missing tuples
needed to form a seamless join. The last figure shows the pattern where
$y-1$ empty blocks separate the two non-empty blocks, and the count sequence
in the second block starts at $s = 1 + (-y \bmod N_2)$. 
}
\end{figure}

To prove that ordering the columns by increasing cardinality minimizes the number
of runs, 
it is enough to prove that permuting the columns two-by-two,
so as to put the column with lesser cardinality first, never increases
the number of runs. To prove this result, we need the following technical lemma.

\begin{lemma}\label{lemma:techlemmalexico1}
For $1\leq N_2< N_3\leq  30$ and $0<p\leq 1$, we have
\begin{eqnarray*}
 (1-P_{\downdownarrows N_3} )  \rho_{N_3} N_2
-P_{\downdownarrows N_2,\rho_{N_3}} \rho_{N_2,\rho_{N_3}}
< (1-P_{\downdownarrows N_2}) \rho_{N_2}   N_{3}
 - P_{\downdownarrows N_3,\rho_{N_2}} \rho_{N_3,\rho_{N_2}}.
 \end{eqnarray*}
\end{lemma}
\begin{proof}
Observe that
$1-\rho_{N_3}=(1-p)^{N_3}$ and 
$\rho_{N_2,\rho_{N_3}}= 1-(1-p)^{N_2 N_3}= \rho_{N_2 N_3}$.
To prove the result, we show that:
\begin{itemize}
\item For $p$ sufficiently close to $1$, the result holds.
\item We can turn the inequality into a polynomial in $p$ with no root in $(0,1)$.
\end{itemize}

The first item is easy: taking the limit as $p\rightarrow 1$ on both sides
of the inequality, we get $N_2<N_3$.
To conclude the proof, we have to show that
$ (1-P_{\downdownarrows N_3} )  \rho_{N_3} N_2
-P_{\downdownarrows N_2,\rho_{N_3}} \rho_{N_2,\rho_{N_3}}
- (1-P_{\downdownarrows N_2}) \rho_{N_2}   N_{3}
 + P_{\downdownarrows N_3,\rho_{N_2}} \rho_{N_3,\rho_{N_2}}$ is never
 zero for $p \in (0,1)$.
 We multiply this quantity by
$\rho_{N_2 N_3}$. We 
proceed  
to show that the result 
is a polynomial.

Since $1-z^N = (1-z) (1+z+\cdots + z^{N-1})$, we have
that  the polynomial $\rho_{N_2 N_3}$ is divisible by both $\rho_{N_2}$ and $\rho_{N_3}$  by respectively setting  $z=(1-p)^{N_2}$ and $z=(1-p)^{N_3}$.
Hence, 
$(1-P_{\downdownarrows N_3} )  \rho_{N_3} \rho_{N_2 N_3}$
and 
$(1-P_{\downdownarrows N_2} )  \rho_{N_2} \rho_{N_2 N_3}$
are polynomials.

We also have that 
$P_{\downdownarrows N_2,\rho_{N_3}} \rho_{N_2,\rho_{N_3}}
= \frac{N_2 \rho_{N_3}^2 (1-\rho_{N_3})^{N_2-1}}{\rho_{N_2 N_3}}$
and similarly for 
$P_{\downdownarrows N_3,\rho_{N_2}} \rho_{N_3,\rho_{N_2}}$
so that
$(P_{\downdownarrows N_2,\rho_{N_3}} \rho_{N_2,\rho_{N_3}}
-P_{\downdownarrows N_3,\rho_{N_2}} \rho_{N_3,\rho_{N_2}})\rho_{N_2 N_3}$
is a polynomial.

Hence, for any given $N_2$ and $N_3$, we can check that the result
holds by applying Sturm's method~\cite{barnard08}
 to the polynomial over the interval
$(0,1]$. Because there is no root at $p=1$, we have to check
that the total root count over $(0,1]$ is always zero.  We proved this result using a computer algebra system
(see Appendix~\ref{appendix:maximacode}) for values of $N_2$ and $N_3$ up to 30.
This concludes the proof.
\end{proof}

There are $N_1-1$~pairs of blocks immediately adjacent,
$N_1-2$~pairs of blocks separated by a single block, and so on.
Hence, the expected number of seamless joins in the second column is\footnote{We 
use the identity $\sum_{k=0}^{N-2} (N-1-k)x^k= \frac{(1-x) N +x^N-1}{(1-x)^2}$.}
$S_{N_1,N_2}^{\mathrm{lexico}}=P_{\downdownarrows N_2} \rho_{N_2}^2 \sum_{k=0}^{N_1-2} (N_1 -1 -k ) (1-\rho_{N_2})^k$
or 
$S_{N_1,N_2}^{\mathrm{lexico}}=P_{\downdownarrows N_2}  ( \rho_{N_2} N_1 +(1-\rho_{N_2})^{N_1}-1 )=P_{\downdownarrows N_2}  \rho_{N_2} N_1+\epsilon$ for $|\epsilon|\leq 1$. 
\begin{proposition}\label{prop:lexico}
Consider a table with $c$ independent and uniformly distributed columns having
cardinalities $N_1,N_2,\ldots,N_c$ (let $2\leq N_i\leq N_{i+1}\leq 30$ for $i=1,\ldots,c-1$). 
We can sort the table by lexicographic order according to various column orders.
The column order  $N_1,N_2,\ldots,N_c$  minimizes the number of column runs---up to a term no larger than $c$ in absolute value.
\end{proposition}
\begin{proof} 
Define 
$T_{N_1, N_2, \rho_{N_3}}^{\mathrm{lexico}} =N_1 N_2 \rho_{N_3} - P_{\downdownarrows N_2,\rho_{N_3}}  \rho_{N_2,\rho_{N_3}} N_1 $
as the number of expected number of runs---up to a constant term no larger than one in absolute value---in the second column of a 3-column table with cardinalities $N_1, N_2, N_3$ and uniform distribution.
Define $T_{N_1 N_2,N_3,p}^{\mathrm{lexico}}$, $T_{N_1,N_3,\rho_{N_2}}^{\mathrm{lexico}}$ and $T_{N_1 N_3,N_2,p}^{\mathrm{lexico}}$ similarly.
It is sufficient to prove that 
$T_{N_1, N_2, \rho_{N_3}}^{\mathrm{lexico}} + T_{N_1 N_2,N_3,p}^{\mathrm{lexico}} 
\leq T_{N_1,N_3,\rho_{N_2}}^{\mathrm{lexico}} 
+  T_{N_1 N_3,N_2,p}^{\mathrm{lexico}}
$ whenever $N_2\leq N_3$,
irrespective of the value of $N_1$ (allowing $N_1>N_3$). 
We have 
\begin{eqnarray*}T_{N_1, N_2, \rho_{N_3}}^{\mathrm{lexico}} + T_{N_1 N_2,N_3,p}^{\mathrm{lexico}}
& = & N_1 N_2 \rho_{N_3} - P_{\downdownarrows N_2,\rho_{N_3}}   \rho_{N_2,\rho_{N_3}} N_1 \\
&& + N_1 N_2 N_3 p - P_{\downdownarrows N_3}  \rho_{N_3} N_1 N_2 \\
& = & (1- P_{\downdownarrows N_3}) \rho_{N_3} N_1 N_2 \\
&& - P_{\downdownarrows N_2,\rho_{N_3}}   \rho_{N_2,\rho_{N_3}} N_1 \\
&& + N_1 N_2 N_3 p \\
& \leq & (1- P_{\downdownarrows N_2}) \rho_{N_2} N_1 N_3 \\
&& - P_{\downdownarrows N_3,\rho_{N_2}}   \rho_{N_3,\rho_{N_2}} N_1 \\
&& + N_1 N_2 N_3 p  ~\textrm{(by Lemma~\ref{lemma:techlemmalexico1})}\\
& =& 
N_1 N_3 \rho_{N_2} - P_{\downdownarrows N_3,\rho_{N_2}}   \rho_{N_3,\rho_{N_2}} N_1 \\
&&+ N_1 N_2 N_3 p - P_{\downdownarrows N_2}  \rho_{N_2} N_1 N_3 \\
& = & 
T_{N_1,N_3,\rho_{N_2}}^{\mathrm{lexico}} 
+  T_{N_1 N_3,N_2,p}^{\mathrm{lexico}}.
\end{eqnarray*}
This proves the result. 
\end{proof}

We conjecture that a similar result would hold for all values of $N_i$ larger
than 30. Given arbitrary values of $N_1, N_2,\ldots, N_c$, we can quickly check
whether the result holds using a computer algebra system.

\subsubsection{Best column order for reflected Gray-code sorting}
\label{sec:ExpectednumberofrunsafterGraycodesorting}

For the reflected Gray-code order, we want to prove that the best column ordering
is by increasing cardinality, irrespective of the column cardinalities.
Blocks in reflected Gray-code sort are either ordered from 1 to $N_2$,
or from $N_2$ to 1. When two non-empty
blocks of the same type are separated by empty blocks,
the probability of having a seamless join is $P_{\downdownarrows N_2}$.
Otherwise, the probability of seamless join is
\label{pud-defined}
  $P_{\updownarrow N_2} = \frac{p^2 + (1-p)^2 p^2 +\cdots+ (1-p)^{2N_2-2} p^2}{(1-(1-p)^{N_2})^2}
= \frac{p^2 (1-(1-p)^{2N_2})}{(1-(1-p)^{N_2})^2 (1-(1-p)^{2})}
$ for $p\in (0,1)$. 

There are $N_1-1$~pairs of blocks immediately adjacent and with opposite orientations (e.g., from 1 to $N_2$ and then from $N_2$ to 1; see 
Fig.~\ref{fig:rgc-seamless}),
$N_1-2$~pairs of blocks separated by a single block and having identical orientations, and so on.
Hence, the expected number of seamless joins is
$P_{\updownarrow N_2} \rho_{N_2}^2 \sum_{k=0}^{\lfloor (N_1-1)/2 \rfloor} (N_1 -1 -2k ) (1-\rho_{N_2})^{2k}  +
P_{\downdownarrows N_2} \rho_{N_2}^2 \sum_{k=0}^{\lfloor (N_1-3)/2 \rfloor} (N_1 -2 -2k ) (1-\rho_{N_2})^{2k+1} 
$. 

We want a simpler formula for the number of runs, at the expense of introducing an error of plus or minus one run. So consider the scenario where we have an infinitely long column, instead of
just $N_{1}$~blocks. However, we count only the number of seamless joins between a block
in the first $N_{1}$~blocks and a block following it. Clearly, there can be at most one extra
seamless join, compared to the number of seamless joins within the  $N_{1}$~blocks.

We have the formula $x \sum_{k=0}^{\infty} (1-x)^{2k} =\frac{x}{1-(1-x)^2} = \frac{1}{2-x}$. 
Hence, this new  number of seamless joins is 
$S_{N_1,N_2}^{\mathrm{reflected}}=
P_{\updownarrow N_2} \rho_{N_2}^2 \sum_{k=0}^{\infty} N_1 (1-\rho_{N_2})^{2k}  +
P_{\downdownarrows N_2} \rho_{N_2}^2 \sum_{k=0}^{\infty} N_1  (1-\rho_{N_2})^{2k+1} =
\frac{P_{\updownarrow N_2} \rho_{N_2} N_{1}}{2-\rho_{N_2} } +
\frac{P_{\downdownarrows N_2} \rho_{N_2}(1-\rho_{N_2}) N_{1}}{2-\rho_{N_2} }.
$

Let $\lambda^{\mathrm{reflected}}_{N_{2}} = 
\frac{ P_{\updownarrow N_2} +(1-\rho_{N_2}) P_{\downdownarrows N_2}}{2-\rho_{N_2} } $,
then 
$S_{N_1,N_2}^{\mathrm{reflected}}=\lambda^{\mathrm{reflected}}_{N_{2}} \rho_{N_2} N_{1}$.

\begin{lemma}\label{lemma:boringandtechnical}
We have that $\frac{1-x^{ N_2 N_3}}{1- x^{ N_3}} = 1+x^{N_3} + x^{2 N_3} + \cdots + x^{N_3 (N_2-1)},$ for all positive integers $N_2, N_3$.
\end{lemma}

\begin{lemma}\label{lemma:techlemmareflected1}
If $2\leq N_{2}< N_{3} \leq 30 $, then  
\begin{eqnarray*} (1-  \lambda^{\mathrm{reflected}}_{N_{3}}) \rho_{N_3}  N_2
-\lambda^{\mathrm{reflected}}_{N_2,\rho_{N_3}} \rho_{N_2,\rho_{N_3}} 
<
(1- \lambda^{\mathrm{reflected}}_{N_2}) \rho_{N_2}  N_3
- \lambda^{\mathrm{reflected}}_{ N_3,\rho_{N_2}}   \rho_{N_3,\rho_{N_2}}.
\end{eqnarray*} 
\end{lemma}
\begin{proof} The proof is similar to the proof of Lemma~\ref{lemma:techlemmalexico1}.
We want to show that:
\begin{itemize}
\item For some value of $p$ in (0,1), 
the result holds.
\item We can turn the inequality into a polynomial in $p$ with no root in $(0,1)$.
\end{itemize}

The first item follows by evaluating the derivative of both sides
of the inequality at $p=1$. (Formally, our formula is defined for $p\in (0,1)$, so
we let the values and derivatives of our functions at 1 be implicitly defined
as their limit as $p$ tends to 1.) For all $N\geq 2$ and  at $p=1$, we have that $\frac{d P_{\updownarrow N}}{dp}=2$, 
$\frac{d P_{\downdownarrows N}}{dp}=0$, $\frac{d \rho_{N}}{dp}=0$, and $\frac{d \lambda^{\mathrm{reflected}}_{N}}{dp}=2$; moreover, we have $\rho_{N}=1$ and $\lambda^{\mathrm{reflected}}_{N}=1$ at $p=1$.
The derivatives of $P_{\updownarrow N,\rho{N'}}$, $P_{\downdownarrows N,\rho{N'}}$
and $\lambda^{\mathrm{reflected}}_{N,\rho{N'}}$ are also zero for all $N,N'\geq 2$ at $p=1$. 
 Hence, the derivative of the left-hand-side of the inequality at $p=1$ is $-2 N_2$ whereas the
  derivative of the right-hand-side is $-2 N_3$. Because $N_3>N_2$ and equality holds at $p=1$,
   we have that the left-hand-side must be smaller than the right-hand-side at $p=1-\xi$
    for some sufficiently small $\xi>0$.

To conclude the proof, we have to show that 
the value 
$ (1-  \lambda^{\mathrm{reflected}}_{N_{3}}) \rho_{N_3}  N_2
- \lambda^{\mathrm{reflected}}_{N_2,\rho_{N_3}} \rho_{N_2,\rho_{N_3}} 
-
(1- \lambda^{\mathrm{reflected}}_{N_2}) \rho_{N_2}  N_3
+ \lambda^{\mathrm{reflected}}_{ N_3,\rho_{N_2}}   \rho_{N_3,\rho_{N_2}}$ is never
 zero for $p \in (0,1)$.
 We multiply this quantity by
$(2- \rho_{N_2 N_3}) \rho_{N_2 N_3}$ and call the result $\Upsilon$. We 
first
show that $\Upsilon$
is a polynomial.

Because $P_{\updownarrow N_3}\rho_{N_3}^2$ and $P_{\downdownarrows N_3}\rho_{N_3}^2$ are polynomials (respectively $p^2 + (1-p)^2 p^2 +\cdots+ (1-p)^{2N_2-2} p^2$ and $N_2 p^2 (1-p)^{N_2-1}$),
we have that $\lambda^{\mathrm{reflected}}_{N_{3}}$ can be written as
a polynomial divided by $(2-\rho_{N_3}) \rho_{N_3}^2$.
Hence, 
$ \lambda^{\mathrm{reflected}}_{N_{3}} \rho_{N_3} (2- \rho_{N_2 N_3}) \rho_{N_2 N_3}$ is a polynomial times $\frac{ (2- \rho_{N_2 N_3}) \rho_{N_2 N_3}}{(2-\rho_{N_3}) \rho_{N_3}}$. In turn, this fraction is $\frac{1-(1-p)^{2 N_2 N_3}}{1-(1-p)^{2 N_3}}$ which is a polynomial by Lemma~\ref{lemma:boringandtechnical}.
Hence, 
$ \lambda^{\mathrm{reflected}}_{N_{3}} \rho_{N_3} (2- \rho_{N_2 N_3}) \rho_{N_2 N_3}$ is a polynomial. By symmetrical arguments,
$ \lambda^{\mathrm{reflected}}_{N_{2}} \rho_{N_2} (2- \rho_{N_2 N_3}) \rho_{N_2 N_3}$ is also a polynomial.

Recall that $\rho_{N_2,\rho_{N_3}}= \rho_{N_2N_3}$. We have
that $\lambda^{\mathrm{reflected}}_{N_2,\rho_{N_3}} $ is a polynomial divided by
$(2-\rho_{N_2 N_3}) \rho_{N_2 N_3}^2$. Hence, it is immediate
that $\lambda^{\mathrm{reflected}}_{N_2,\rho_{N_3}} \rho_{N_2,\rho_{N_3}} $ multiplied by $(2- \rho_{N_2 N_3}) \rho_{N_2 N_3}$ is polynomial, merely by canceling the terms in the denominator. A symmetrical argument applies to  
$\lambda^{\mathrm{reflected}}_{N_3,\rho_{N_2}} \rho_{N_3,\rho_{N_2}} $.

Hence, 
 $\Upsilon$ is a polynomial.
As in Lemma~\ref{lemma:techlemmalexico1}, for any given $N_2$ and $N_3$, we can check that there are no roots by applying Sturm's method to the polynomial over the interval
$(0,1]$. Because there is a root at $p=1$, it is sufficient to check
that the total root count over $(0,1]$ is always one.
(Alternatively, we could first divide the polynomial by $x-1$ and check that
there is no root.)
 We proved this result using a computer algebra system
(see Appendix~\ref{appendix:maximacode}).
This concludes the proof.
\end{proof}

\begin{proposition}\label{prop:graycode}
Consider a table with $c$ independent and uniformly distributed columns having
cardinalities $N_1,N_2,\ldots,N_c$ (let $2\leq N_i\leq  N_{i+1}\leq 30$ for $i=1,\ldots,c-1$). 
We can sort the table by reflected Gray-code order according to various column orders.
The column order  $N_1,N_2,\ldots,N_c$ minimizes the number of column runs---up to
a term no larger than $c$ in absolute value.
\end{proposition}


\begin{proof} The proof is similar to Proposition~\ref{prop:lexico}, see Appendix~\ref{appendix:proofofgraycode}.
\end{proof}

\section{\uppercase{Experiments}}
\label{sec:experiments}

To complete the mathematical analysis, we ran experiments on realistic data sets.
We are motivated by the following questions:
\begin{itemize}
\item For columns with few columns, is recursive sorting nearly optimal? (\S~\ref{sec:recursivesortingisafe})
\item How likely is it that
alternative column order are preferable to the increasing-cardinality order? (\S~\ref{sec:isheuristicreliable})
\item How significant can the effect of the column order be? 
 Are reflected Gray-code orders better than lexicographical orders? (\S~\ref{sec:columnordermatters})
 \item How does an Hilbert order compare to lexicographical orders? (\S~\ref{sec:hilbertnotgood})
 \item How large is the effect of skew and column dependency? (\S~\ref{sec:skewanddepend})
 \item Do our results extend to other column-compression techniques? (\S~\ref{sec:effectoncompressiontechniques})
 \end{itemize}

\subsection{Software}

We implemented the various sorting techniques using Java
and the Unix command \texttt{sort}. For all but lexicographic ordering,
hexadecimal values were prepended to each line in a preliminary pass
over the data, before the command \texttt{sort} was called.
(This approach is recommended by Richards~\cite{richards1986dca}.)
Beside recursive orders, we also implemented sorting by Compact Hilbert Indexes (henceforth Hilbert)~\cite{hamilton2007chi}---also by prepending hexadecimal values.
By default, we order values within columns alphabetically.

\subsection{Realistic data sets}

We used five data sets (see Table~\ref{tab:caractDataSet})
representative of tables found in applications: Census-Income~\cite{KDDRepository}, Census1881~\cite{prdh:website}, DBGEN~\cite{DBGEN}, Netflix~\cite{netflixprize} and KJV-4grams~\cite{arxiv:0901.3751}.
The Census-Income table has 4~columns: \textit{age}, 
\textit{wage per hour}, \textit{dividends from stocks} 
and a numerical value\footnote{%
The associated metadata says this column should be a
10-valued migration code.} found in the $25^{\mathrm{th}}$~position of the original data set. The respective cardinalities 
are 91, 1\,240, 1\,478 and 99\,800.
The Census1881 came from a publicly available
 SPSS file 1881\_sept2008\_SPSS.rar~\cite{prdh:website}
that we converted to a flat file. In the process, we replaced the special
values ``ditto'' and ``do.\@'' by the repeated value, and we
deleted all commas within values.
The column cardinalities are 183, 2\,127, 2\,795, 8\,837, 24\,278, 152\,365, 152\,882.
For DBGEN, we selected dimensions of cardinality 7, 11, 2\,526 and 400\,000.
The Netflix  table has 
4~dimensions:  UserID, MovieID, Date and Rating,
with cardinalities 
480\,189, 17\,770, 2\,182, and 5.
Each of the four columns of KJV-4grams contains roughly 8~thousand distinct stemmed words:
8\,246, 8\,387, 8\,416, and 8\,504.

Table~\ref{tab:caractDataSet} also gives the suboptimality factor $\mu$ from Proposition~\ref{prop:optimalitybound}.
For DBGEN, any recursive order minimizes the number of runs optimally---up to a factor of 1\%.
For Netflix and KJV-4-grams, recursive ordering is 2-optimal. Only for Census1881 is the bound on
optimality significantly weaker: in this instance, recursive ordering is 5-optimal.

\begin{table}\small
     \caption{Characteristics of data sets used
    }\label{tab:caractDataSet}
    \centering
    \begin{tabular}{l|rrrrrr|} \cline{2-7}
     & rows & distinct rows & cols & \rule{0mm}{1.1em} $\sum_i n_i$ & size & $\mu$ \\ \hline
    \multicolumn{1}{|l|} {\textbf{Census-Income} }        & 199\,523 & 178\,867    & 4 & 102\,609   & 2.96\,MB &2.63  \\
    \multicolumn{1}{|l|}{\textbf{Census1881} }         & 4\,277\,807 & 4\,262\,238 & 7 & 343\,422   & 305\,MB &5.09 \\
    \multicolumn{1}{|l|}{\textbf{DBGEN} }         & 13\,977\,980 & 11\,996\,774 & 4 & 402\,544   & 297\,MB &1.02 \\
    \multicolumn{1}{|l|}{\textbf{Netflix}}       & 100\,480\,507 &100\,480\,507 & 4 & 500\,146   & 2.61\,GB &2.00\\
    \multicolumn{1}{|l|}{\textbf{KJV-4grams}}    & 877\,020\,839 & 363\,412\,308 & 4 &  33\,553   & 21.6\,GB  &2.19\\ 
\hline
    \end{tabular}

\end{table}

\subsection{Recursive sorting is ``safe'' for low dimensionality}
\label{sec:recursivesortingisafe}
Since our 7-dimensional data set yields a much looser bound than the 4-dimensional data sets, 
we investigate the relationship between $\mu$ and the number of dimensions.
Rather than use our arbitrarily chosen low-dimensional projections, we randomly generated
many projections (typically 1000) of each original data set, computed $\mu$ for each projection, then showed the $\mu$
values for each dimensionality (i.e., all $\mu$ values for 3-dimensional projections
were averaged and reported; likewise all $\mu$ values for 4-dimensional projects were averaged
and reported). One difficulty arose: computing $\mu$ for a projection 
required the number of distinct rows, and we projected from data sets that are at least a
large fraction of our main-memory size.  Gathering this data exactly appears too expensive. 
Instead, we computed the projection sizes
in a few passes over our full data sets, using a probabilistic counting
technique due to Cai~et~al.~\cite{cai2005fat} that was shown by 
Aouiche and Lemire~\cite{aouiche2007cfp} to have a good performance.
As an extra step, we corrected the
distinct-row estimates so that they never exceeded the product of column cardinalities.
 To validate our estimates,
  we computed exact $\mu$ values
for two smaller data sets 
(TWEED~\cite{tweed,diamond-ideas} with 52 dimensions and 11k rows,
and another with 13 dimensions and 581k rows) and observed our average
$\mu$ estimates changed by less than 2\%.
KJV-4grams and Netflix only had 4 dimensions,
and thus we used TWEED to get another high-dimensional
data set.

 \begin{figure}
\subfloat[\textsc{Realistic data sets}\label{fig:mu-vs-d}]{
\includegraphics[width=0.49\textwidth]{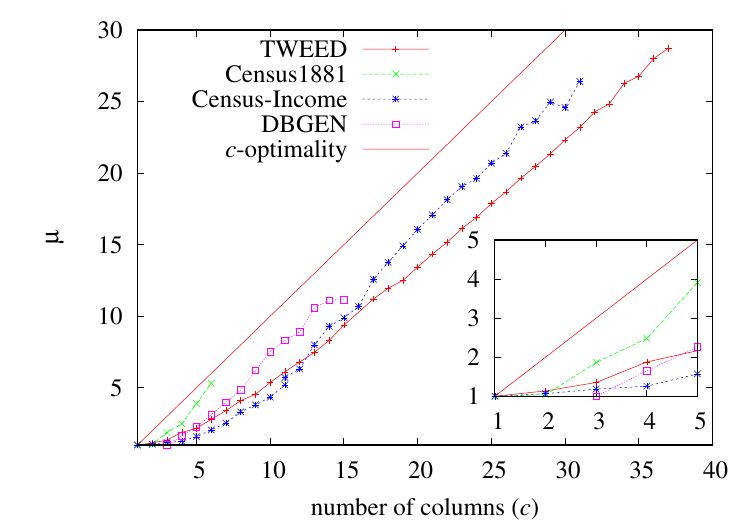}
}
\subfloat[\textsc{Synthetic data sets}\label{synthfig:mu-vs-d}]{
\includegraphics[width=0.49\textwidth]{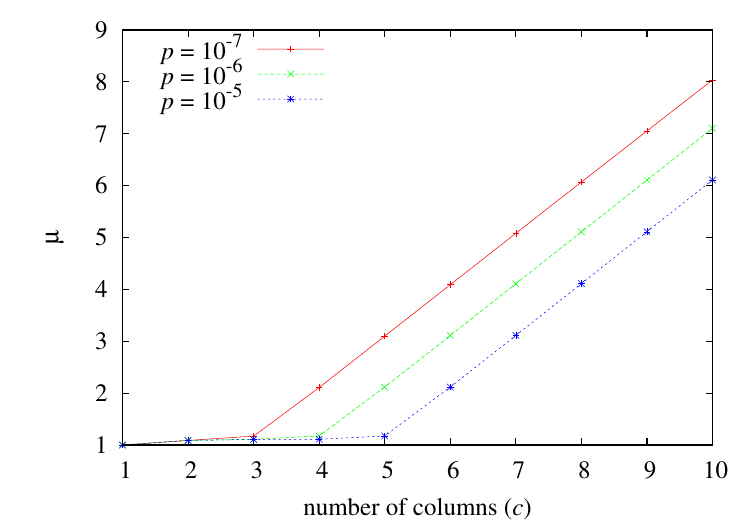}
}
\caption{Approximate $\mu$ versus columns, when sampling projections of
realistic data sets and synthetic data sets (10-dimensional uniformly distributed table with $N_1=N_2=\dots=N_{10}=10$)}
 \end{figure}


Figure~\ref{fig:mu-vs-d} shows that, after about 3 dimensions, 
$\mu$ grew roughly linearly with 
the number of dimensions: the $\mu$ 
formula's $\min(n,N_{1,i})/(n+c-1)$ terms would
typically approximate 1 for all but the first few dimensions.
To illustrate this
, we computed the expected value of $\mu$
for projections of uniformly distributed tables with various densities $p$ (see Fig.~\ref{synthfig:mu-vs-d}).

%
This does \emph{not} mean that any particular recursive sorting
algorithm will be this far from optimal.  Our $\mu$ is an upper bound on suboptimality, 
so it merely means that we have not given evidence
that recursive sorting is necessarily good for higher-dimension data sets.
For high-dimensional data sets, there could still be a significant advantage
in going beyond lexicographic sorting or other recursive sorting approaches.
However, for 2 or 3 dimensions, our $\mu$ values show that lexicographic
sorting cannot be improved much.  Of course, such projections may 
be nearly complete tables (cf.~\S~\ref{sec:complete-tables}).

\subsection{The column-reordering heuristic is reliable}
\label{sec:isheuristicreliable}

\begin{figure}
\centering
\includegraphics[width=0.6 \columnwidth]{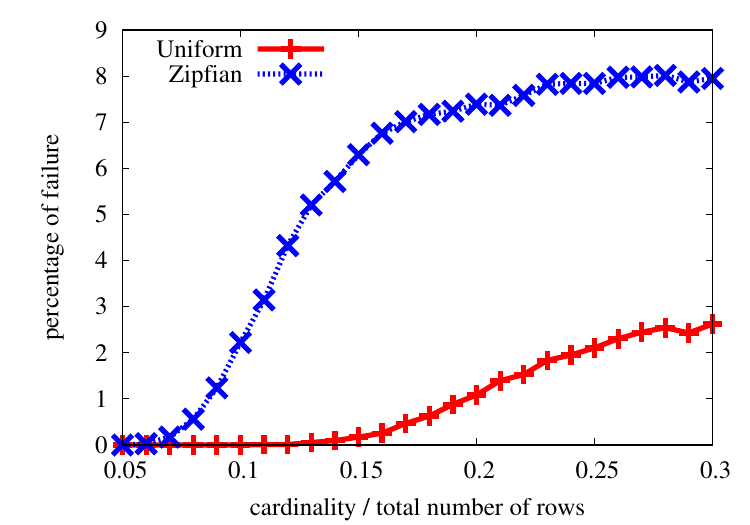}
\caption{\label{fig:CrazyExperimentToSatisfyOwen}Percentage of failure of the increasing-cardinality column-reordering heuristic }
\end{figure}

We showed in \S~\ref{sec:Increasing-cardinality-order} that
reordering the columns in increasing cardinality minimized the
expected number of runs. To assess the reliability 
of this heuristic, consider a two-dimensional table model 
where the first column's values are selected
uniformly at random from 1 to $N_1$, and the
second column's values are selected uniformly from 1 to $N_1+1$:
we want the second column to have just barely a higher cardinality 
than the first.
Using this model, we generated 100\,000 100-row tables for each 
column cardinality $N_1$ from 5 to 30. 
Of course, we can expect a few missing values when selecting
100 items (with replacement) from $N_1$.
Some tables had more missing values in the second
column, so we kept only the  randomly generated
tables where the second column had the higher actual cardinality.
We then determined the percentage of
tables where the increasing-cardinality column-reordering heuristic failed
to be at least as good as the alternative column order (see Fig.~\ref{fig:CrazyExperimentToSatisfyOwen}).
The expected relative difference between the cardinalities  ranges from  $\approx 1/5$ to $\approx 1/30$.
We see that the rate of failure increases as the relative difference between the column
cardinalities goes to zero. Even so,
the rate of failure is relatively low in this test (less than 3\%) despite the small
relative difference in cardinality. 

Although our theoretical results assume uniformity, we have observed that 
ordering columns
in ascending order also tends to improve results with skewed data.
To assess reliability, we repeated the same test for Zipfian-distributed columns and found the rate of failure was larger. However, it still remained moderate (less than 9\%).
Moreover, even with Zipfian distributions, the rate of failure is close to zero when the
relative difference between the column cardinalities is large (0.05).

\subsection{Column order matters, Gray codes do not }
\label{sec:columnordermatters}

Results for realistic data sets are given in Table~\ref{table:mainresults}. 
For these data sets, there is no noticeable benefit (within 1\%) to Gray codes as opposed to lexicographic orders.  The only data set showing some benefit ($\approx 1$\%) is KJV-4grams.

\begin{table}
\caption{\textsc{RunCount} after sorting various tables using different orderings. The up  and down arrows indicate whether the columns where ordered in increasing or decreasing cardinality before sorting. Best results for each data set are in bold.\label{table:mainresults}
 }
\centering
	\begin{tabular}{cccccc}\toprule
table                      & shuffled    & order&     lexico.          & Gray &Hilbert \\  \midrule
\multirow{2}{*}{Census-Income}& \multirow{2}{*}{4.6$\times 10^5$} &  $\downarrow$  & 3.2$\times 10^5$ & 3.2$\times 10^5$ & 3.4$\times 10^5$ \\
  &  &$\uparrow$ & \bm{$1.9\times 10^5$} & \bm{$1.9\times 10^5$} & 3.4$\times 10^5$ \\\midrule[0.1pt]
\multirow{2}{*}{Census1881} & \multirow{2}{*}{2.7$\times 10^7$} &$\downarrow$  &  1.8$\times 10^7$&  1.8$\times 10^7$&  2.0$\times 10^7$ \\ 
  &  &$\uparrow$ &  \bm{$1.3\times 10^7$}&  \bm{$1.3\times 10^7$}&  2.0$\times 10^7$ \\\midrule[0.1pt]
\multirow{2}{*}{DBGEN}  & \multirow{2}{*}{4.5$\times 10^7$} & $\downarrow$ & 3.3$\times 10^7$ & 3.3$\times 10^7$ &  4.3$\times 10^7$ \\
 &   &$\uparrow$& \bm{$1.2\times 10^7$} &\bm{$1.2\times 10^7$} & 4.3$\times 10^7$ \\\midrule[0.1pt]
\multirow{2}{*}{Netflix} & \multirow{2}{*}{3.8$\times 10^8$} &  $\downarrow$& 2.5$\times 10^8$ & 2.5$\times 10^8$ & 3.3$\times 10^8$ \\
  &   &$\uparrow$ & \bm{$1.2\times 10^8$} & \bm{$1.2\times 10^8$} & 3.3$\times 10^8$ \\\midrule[0.1pt]
\multirow{2}{*}{KJV-4grams}  & \multirow{2}{*}{3.4$\times 10^9$} & $\downarrow$& 3.9$\times 10^8$ & \bm{$3.8\times 10^8$} & 8.2$\times 10^8$ \\
 &  &$\uparrow$ & 3.9$\times 10^8$ & \bm{$3.8\times 10^8$} & 8.2$\times 10^8$ \\
 \bottomrule
	\end{tabular}
\end{table}


Relative to the shuffled case, 
ordering the columns in increasing cardinality reduced the number of runs by a factor of two (Census and Census1881), three (DBGEN and Netflix) or nine (KJV-4grams). Except for Netflix and KJV-4grams, these gains drop to  $\approx 50$\% when using the wrong column order (by decreasing cardinality). On Netflix, the difference between the two column orders is a factor of two (2.5$\times 10^8$ versus 1.2$\times 10^8$).
 
 The data set KJV-4grams appears oblivious to column reordering. We are not surprised
 given that columns have similar
 cardinalities and distributions.


%
%

\subsection{Compact Hilbert Indexes are not competitive}
\label{sec:hilbertnotgood}

Hilbert is effective at improving the compression of 
database tables~\cite{eavis2007hilbert,dehne2007compressing} using 
tuple difference coding techniques~\cite{ng1997block}.
Moreover, for complete tables where the cardinality of all columns
is the same power of two, sorting by Hilbert  minimizes
the number of runs (being a Gray code \S~\ref{sec:lex-and-gc-sorting}).
However, we are unaware of any application of Hilbert to 
column-oriented indexes.

To test Hilbert, we generated a small random table (see Table~\ref{table:comparinghilbertwithothers}) with moderately low density ($p=0.01$).
The \textsc{RunCount} result is far worse than recursive ordering, even when all column cardinalities are the same power of 2. 
In this test, Hilbert is column-order oblivious. 
We have similar results over realistic data sets (see Table~\ref{table:mainresults}).
In some instances, Hilbert is nearly as bad as a random shuffle of the table, and always
inferior to a mere lexicographic sort.
For KJV-4grams, Hilbert is relatively effective---reducing the number of runs by a factor of 4---but it is still half as effective as lexicographically sorting the data.

\begin{table}
\caption{Comparison of Compact Hilbert Indexes with other orderings for a uniformly distributed table ($p=0.01$, $c=5$) and various column cardinalities. The number
of runs is given in thousands.\label{table:comparinghilbertwithothers}
}
\centering
\begin{tabular}{cccccc}
\hline
cardinalities & shuffled & lexico. & reflected Gray & modular Gray & Hilbert
\\
\hline
4,8,16,32,64 & 47.8 & 18.9 & \textbf{18.7} & 18.7 & 35.3 \\
64,32,16,8,4 & 47.8 & 28.5 & \textbf{28.1} & 28.2 & 35.3 \\
16,16,16,16,16 & 49.7 & 23.7 & \textbf{23.3} & 23.4 & 35.3 \\ \hline
\end{tabular}
\end{table}

\subsection{The order of 
values is irrelevant}

For several recursive orders (lexicographic and Gray codes), we 
reordered the attribute values by their frequency---putting the most frequent values first~\cite{Kaser20062304}. 
While the number of runs in sorted uniformly distributed tables is oblivious  to the order of attribute values, we may see some benefits with tables having skewed distributions. However, on the realistic data sets, the differences were small---less than 1\% on all metrics for recursive ordering.

For Hilbert, reordering the attribute values had small and inconsistent effects. For Census, the number of runs went up from 3.4$\times 10^5$ to 3.6$\times 10^5$ (+6\%), whereas for Netflix, it went 
down from 3.3$\times 10^8$ to 3.2$\times 10^8$ (-3\%). The strongest effect was
observed with KJV-4grams where the number of runs went down from 8.2$\times 10^8$ to 7.6$\times 10^8$ (-7\%). These differences are never sufficient to make Hilbert competitive.



%

%

\subsection{Skew and column dependencies reduce the number of runs}
\label{sec:skewanddepend}

\begin{figure}
\caption{Number of runs for the Census-Income data set, including the expected
number assuming that the table is uniformly distributed. The columns are indexed from 1 to 4 in increasing cardinality. Hence, the label 1234 means that the columns are in increasing cardinality.}\label{fig:compareallpossiblereorderings}
\centering\includegraphics[height=0.45\textwidth,angle=0]{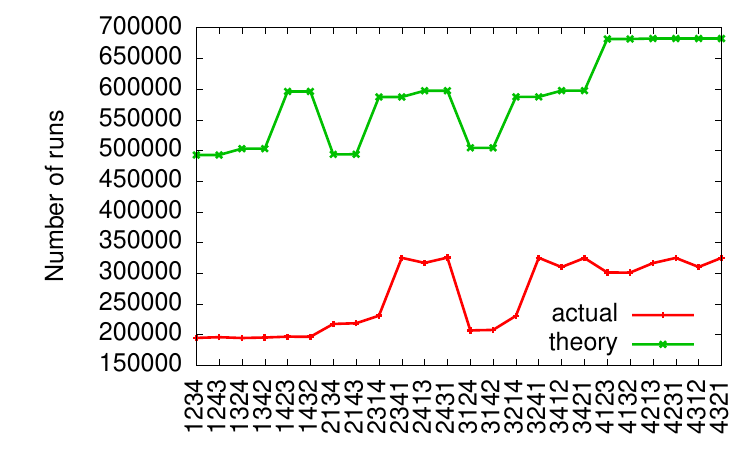}
 \end{figure}

We can compute the expected number of runs for uniformly distributed tables
sorted lexicographically by
the proof of Proposition~\ref{prop:lexico}. 
For Census-Income, we compared this value with the  number of runs
for all possible
 column orders (see Fig.~\ref{fig:compareallpossiblereorderings}).
Distribution skew and dependencies between columns make a substantial difference:
the number of runs would be twice as high were Census-Income uniformly distributed
with independent columns. 
 
\subsection{Effect of column order on alternative column-compression techniques}
\label{sec:effectoncompressiontechniques}

For some implementations of RLE, the compressed size of the columns is proportional to the \textsc{RunCount}.
Thus, the \textsc{RunCount} reduction translate directly into proportionally smaller tables and faster decompression.
But what about other compression techniques?
To answer this question, we implemented three column-compression schemes from the SAP~NetWeaver 
platform~\cite{springerlink:10.1007/978-3-642-15105-7_10}: Prefix, Sparse and Indirect Coding.
We found that for a given compression scheme, the decompression time is roughly proportional to the 
compressed size.  
 Thus, we only report the
compressed sizes (see Table~\ref{table:realcompression}). The benefits
of ordering the columns in increasing cardinality can be large (up to a factor of two).
However, there are also instances where ordering the columns in decreasing cardinality is slightly better
(by up to 5\%). Overall, our recommendation to order the columns in increasing cardinality remains valid.

\begin{table}
\caption{Compressed sizes  under different compression schemes (in MB).  The up  and down arrows indicate whether the columns were ordered in increasing or decreasing cardinality before sorting. \label{table:realcompression}
 }
\centering 
\subfloat[Sparse Coding]{
	\begin{tabular}{cccc}\toprule
table                      & shuffled    & order&     lexico.          \\  \midrule
\multirow{2}{*}{Census-Income}& \multirow{2}{*}{ 0.18} &  $\downarrow$       & 0.17  \\
                                                     &  &$\uparrow$          &	\textbf{0.14} \\\midrule[0.1pt]
\multirow{2}{*}{Census1881} & \multirow{2}{*}{11.2} &$\downarrow$                & \textbf{8.3} \\ 
                                             &  &$\uparrow$                  &  	8.8\\\midrule[0.1pt]
\multirow{2}{*}{DBGEN}  & \multirow{2}{*}{14.6} & $\downarrow$   & 12.9 \\
                                                             &   &$\uparrow$  & 	\textbf{10.3} \\
 \bottomrule
	\end{tabular}
}

\subfloat[Indirect Coding]{
	\begin{tabular}{cccc}\toprule
table                      & shuffled    & order&     lexico.          \\  \midrule
\multirow{2}{*}{Census-Income}& \multirow{2}{*}{ 0.26} &  $\downarrow$       & 0.20  \\
                                                     &  &$\uparrow$          & 	\textbf{0.15} \\\midrule[0.1pt]
\multirow{2}{*}{Census1881} & \multirow{2}{*}{12.6} &$\downarrow$                & 8.3 \\ 
                                             &  &$\uparrow$                  &  	\textbf{7.9}\\\midrule[0.1pt]
\multirow{2}{*}{DBGEN}  & \multirow{2}{*}{18.9} & $\downarrow$   & 10.9 \\
                                                             &   &$\uparrow$  & 	\textbf{10.3} \\
 \bottomrule
	\end{tabular}
}

\subfloat[Prefix Coding]{
	\begin{tabular}{cccc}\toprule
table                      & shuffled    & order&     lexico.          \\  \midrule
\multirow{2}{*}{Census-Income}& \multirow{2}{*}{0.27} &  $\downarrow$       & 0.26  \\
                                                     &  &$\uparrow$          & 	\textbf{0.12} \\\midrule[0.1pt]
\multirow{2}{*}{Census1881} & \multirow{2}{*}{12.6} &$\downarrow$                & 	\textbf{ 10.1} \\ 
                                             &  &$\uparrow$                  &  10.3\\\midrule[0.1pt]
\multirow{2}{*}{DBGEN}  & \multirow{2}{*}{13.9} & $\downarrow$   & 13.1 \\
                                                             &   &$\uparrow$  & 	\textbf{10.1} \\
 \bottomrule
	\end{tabular}
}

\end{table}
\section*{\uppercase{Conclusion}}

Unsurprisingly, an effective heuristic to minimize the number of runs and
column-oriented index sizes is to sort lexicographically after reordering the 
columns in increasing cardinality. This heuristic is often recommended~\cite{arxiv:0901.3751,1142548}.
However, 
our results stress the importance of reordering the columns. Picking the wrong
column order can result in a moderate reduction of the number of runs (50\%) whereas
a large reduction is possible (2--3$\times$) when using the right column order.

The benefit of recursive Gray-code orders over lexicographic orders 
is small. Sorting the values within columns has also small effects ($\leq 1\%$)
for several recursive orders. 


\section*{\uppercase{Future Work}}

The first step toward the estimation of the size 
of column indexes
under sorting is to assume that columns are statistically independent. However,
it might possible to lift this assumption by modeling the dependency between columns~\cite{poosala1997sew,1314854}. 

From a practical point of view, we found that sorting tables lexicographically is effective, especially when the columns are ordered in increasing cardinality. Nevertheless,
we  might sometimes want to spend more time reordering rows, even for modest gains.
Thus, we are currently investigating more expensive
row reordering techniques~\cite{Schaller:1999:RHE:646498.831518,Johnson:2004:CLB:1316689.1316693}.

\section*{Acknowledgments}
This work is supported by NSERC grants 155967 and 261437.

\bibliographystyle{elsart-num-sort} 
\bibliography{RLEwithSorting}
\appendix 

\section{Table of Notation}
\label{sec:notation}

\newcommand{\opthline}{}

\begin{tabular}{llll}\hline
Notation & explanation & defined& used in \\ \hline
\opthline
$r_i$    & runs in column $i$        & p.~\pageref{ri-defined}            & \S~\ref{sec:intro}\\[0.5cm]
\opthline
$c$    & number of columns        & p.~\pageref{ri-defined}            & throughout\\[0.5cm]
$n$    & number of rows & p.~\pageref{Ni-defined}            & throughout \\[0.5cm]
$N_i$    & cardinality of column $i$ & p.~\pageref{Ni-defined}            & throughout \\[0.5cm]
$N_{i,j}$    & $\prod_{k=i}^{j} N_k$        & p.~\pageref{defn-n1c}            & throughout\\[0.5cm]
\opthline
$\mu$ & \parbox{.4\textwidth}{recursive sorting is $\mu$-optimal for the run minimization problem} & p.~\pageref{mu-defined} & throughout\\[0.5cm]
\opthline
$\rho_{N_i}$ & \parbox{.4\textwidth}{probability that a block of $N_i$~tuples is nonempty} & p.~\pageref{rho-defined}& \S~\ref{sec:uniformlydistributed} 
\\[0.5cm]
\opthline
$\rho_{N_i,p'}$ & \parbox{.4\textwidth}{same except individual tuples present with probability $p'$ rather than default $p$} &  & \S~\ref{sec:uniformlydistributed} 
\\[1cm]
\opthline
$P_{\downdownarrows N_2}$ & \parbox{.4\textwidth}{with lexicographic sorting, probability that two nonempty
blocks in column 2 seamlessly join} & p.~\pageref{pdd-defined} & \S~\ref{sec:uniformlydistributed} 
\\[1cm] \opthline
$P_{\downdownarrows N_2,p'}$ & \parbox{.4\textwidth}{same except individual
tuples present with probability $p'$ } &  & \S~\ref{sec:uniformlydistributed} 
\\[1cm]
\opthline
$P_{\updownarrow N_2}$ & \parbox{.4\textwidth}{with reflected Gray sorting, probability that two nonempty
blocks in column 2 seamlessly join} & p.~\pageref{pud-defined} & \S~\ref{sec:uniformlydistributed} 
\\[1cm] \opthline
$P_{\updownarrow N_2,p'}$ & \parbox{.4\textwidth}{same except individual
tuples present with probability $p'$ } &  & \S~\ref{sec:uniformlydistributed} 
\\

\hline

\end{tabular}

\section{Maxima Computer Algebra System code}
\label{appendix:maximacode}

For completing some of the proofs, we used Maxima version~5.12.0~\cite{Maxima}. Scripts ran during
about 49~hours on a Mac~Pro  with two double-core Intel Xeon processors (2.66\,GHz) 
and 2\,GiB of RAM.

The proof of Lemma~\ref{lemma:techlemmalexico1} uses the following code which ran for 185~minutes:
\begin{verbatim}
r(N2,p):=1-(1-p)**N2;
Pdd(N2,p):=N2*p**2*(1-r(N2,p))/((1-p)*r(N2,p)**2);
P:(1-Pdd(N3,p))*r(N3,p)*N2- (1-Pdd(N2,p))*r(N2,p)*N3
-Pdd(N2,r(N3,p))*r(N2*N3,p)+Pdd(N3,r(N2,p))*r(N2*N3,p);
P2:P*r(N2*N3,p);

 for n2:2 unless n2>30 do
    (display(n2),   
    for n3:n2+1 unless n3>100 do
     ( nr: nroots(factor(subst([N2=n2,N3=n3],P2)),0,1),
     if(not(nr=0)) then display("ERROR",n2,n3,nr)));
\end{verbatim}

The proof of Lemma~\ref{lemma:techlemmareflected1} uses this code which ran for 46~hours:
\begin{verbatim}
r(N2,p):=1-(1-p)**N2;
Pdd(N2,p):=N2*p**2*(1-r(N2,p))/((1-p)*r(N2,p)**2);
Pud(N2,p):=p**2*(2-r(N2,p))/(r(N2,p)*(1-(1-p)**2));
Lambda(N2,p):=(Pud(N2,p)+(1-r(N2,p))*Pdd(N2,p))/(2-r(N2,p));
P:(1-Lambda(N3,p))*r(N3,p)*N2- (1-Lambda(N2,p))*r(N2,p)*N3
-Lambda(N2,r(N3,p))*r(N2*N3,p)+Lambda(N3,r(N2,p))*r(N2*N3,p);
P2:P*(2-r(N2*N3,p))*r(N2*N3,p);

 for n2:2 unless n2>30 do
    (display(n2),   
    for n3:n2+1 unless n3>100 do
     ( nr: nroots(factor(subst([N2=n2,N3=n3],P2)),0,1),
     if(not(nr=1)) then display("ERROR",n2,n3,nr)));
\end{verbatim}

\section{Proof of Proposition~\ref{prop:graycode}}
\label{appendix:proofofgraycode}

\begin{proof}
Define 
$T_{N_1, N_2, \rho_{N_3}}^{\mathrm{reflected}} =N_1 N_2 \rho_{N_3} - S_{N_1,N_2,\rho_{N_3}}^{\mathrm{reflected}}$
where $S_{N_1,N_2,\rho_{N_3}}^{\mathrm{reflected}}$ is defined as 
$S_{N_1,N_2}^{\mathrm{reflected}}$ after substituting $\rho_{N_3}$ for $p$.
Define $\lambda^{\mathrm{reflected}}_{N_{2},\rho_{N_3}}$, $T_{N_1 N_2,N_3,p}^{\mathrm{reflected}}$, $T_{N_1,N_3,\rho_{N_2}}^{\mathrm{reflected}}$ and $T_{N_1 N_3,N_2,p}^{\mathrm{reflected}}$ similarly.
As in the proof of Proposition~\ref{prop:lexico},
it is sufficient to prove that 
$T_{N_1, N_2, \rho_{N_3}}^{\mathrm{reflected}} + T_{N_1 N_2,N_3,p}^{\mathrm{reflected}} \leq T_{N_1,N_3,\rho_{N_2}}^{\mathrm{reflected}} 
+  T_{N_1 N_3,N_2,p}^{\mathrm{reflected}}
$ whenever $N_2\leq N_3$,
irrespective of the value of $N_1$ (allowing $N_1>N_3$).
 
We have 
\begin{eqnarray*}T_{N_1, N_2, \rho_{N_3}}^{\mathrm{reflected}} + T_{N_1 N_2,N_3,p}^{\mathrm{reflected}}
& = & N_1 N_2 \rho_{N_3} 
-\lambda^{\mathrm{reflected}}_{N_{2},\rho_{N_3}} \rho_{N_2,\rho_{N_3}} N_{1}\\
&& + N_1 N_2 N_3 p - \lambda^{\mathrm{reflected}}_{N_{3}}   \rho_{N_3} N_1 N_2 \\
& = & (1-  \lambda^{\mathrm{reflected}}_{N_{3}}) \rho_{N_3} N_1 N_2 \\
&& - \lambda^{\mathrm{reflected}}_{N_{2},\rho_{N_3}} \rho_{N_2,\rho_{N_3}} N_{1}\\
&& + N_1 N_2 N_3 p\\
& \leq & (1- \lambda^{\mathrm{reflected}}_{ N_2}) \rho_{N_2} N_1 N_3 \\
&& - \lambda^{\mathrm{reflected}}_{ N_3,\rho_{N_2}}   \rho_{N_3,\rho_{N_2}} N_1 \\
&& + N_1 N_2 N_3 p + \lambda^{\mathrm{reflected}}_{N_2} ~\textrm{(by Lemma~\ref{lemma:techlemmareflected1})}\\
& =& 
N_1 N_3 \rho_{N_2} - \lambda^{\mathrm{reflected}}_{N_3,\rho_{N_2}}  \rho_{N_3,\rho_{N_2}} N_1 \\
&&+ N_1 N_2 N_3 p - \lambda^{\mathrm{reflected}}_{N_2}   \rho_{N_2} N_1 N_3\\
& = & 
T_{N_1,N_3,\rho_{N_2}}^{\mathrm{reflected}} 
+  T_{N_1 N_3,N_2,p}^{\mathrm{reflected}}.
\end{eqnarray*}
This proves the result. 
\end{proof}

\section{A Related NP-Completeness Result}
\label{sec:COMLR}

In \S~\ref{sec:colr} we showed it is NP-hard to order columns so as to
minimize the \textsc{RunCount} value after lexicographic sorting.
We now show a related problem is NP-complete.

\paragraph{Column-Ordering-for-Minimax Lexicographic Runcount (COMLR)}
Given a table $T$, an ordering on the values found in each
column, and an integer $K$, is it possible to reorder the columns
of the table, such that when the reordered table is lexicographically
sorted, no column has more than $K$~runs?

\begin{proposition}
COMLR is NP-complete.
\end{proposition}
\begin{proof}
Membership in NP is obvious.  We show COMLR is NP-hard by reduction from 
3SAT~\cite[LO2]{gare:gandj}.
Suppose our 3SAT instance has variables $v_1$ to $v_{|V|}$ and
clauses $C_1$ to $C_m$.  We assume that no clause contains both a
variable and its negation because such a clause can be removed without
affecting satisfiability. 

For every variable $v_i$, the COMLR instance has  three values that can appear in 
tables: $w_i$,  $\bar{w}_i$ and $0_{w_i}$.   They are ordered: $w_i < \bar{w}_i < 0_{w_i}$.    
Moreover, for $a \in \{w_i,\bar{w}_i, 0_{w_i}\}$, 
$b \in \{w_j,\bar{w}_j, 0_{w_j}\}$ and $i \neq j$,
we have $a < b$ if and only if $i<j$.

Two other values are used in the
table, $+\infty$ and $-\infty$ whose orderings with respect to the other values
are as expected.

We construct a table $T$, with $3|V|+2$ rows,
and with a column for each possible literal and a column
for each clause. Hence $T$ has $2|V|+m$ columns.  We describe the columns 
from left to right, beginning with the columns for $\bar{v}_1$ and $v_1$.
See Fig.~\ref{fig:COMLR}.

Consider the literal column associated with $\bar{v}_1$.   It begins
with a run of length $3\times 1 -2$ with the $-\infty$ value.  It then contains
$\bar{w}_1, \bar{w}_1, 0_{w_1}$.  The remainder of the column is
composed of $+\infty$. The next column is for $v_1$.  It begins and ends similarly, but in
the middle it has $w_1, 0_{w_1}, w_1$.
The pairs of columns for the remaining variables then follow.  The column for 
$\bar{v}_i$ begins with a run containing $3i-2$ copies of the $-\infty$ value, then has
$\bar{w}_i, \bar{w}_i  0_{w_i}$, whereas the column for $v_i$ has $w_i, 0_{w_i}, w_i$ between
the run of $-\infty$ and the run of $+\infty$.
Thus, the left part of the table has blocks of size $3\times 2$ arranged
diagonally .
Above the diagonal, we have $-\infty$;  below the diagonal, we have
$+\infty$. (Except that there is a row of $-\infty$ above
everything and a row of $+\infty$ below everything.)

To complete the construction, we have one column per clause.
Consider a clause $\{ l_i, l_j, l_k\}$
where $l_i = v_i$ or $l_i = \bar{v}_i$ and similarly for $l_j$ and $l_k$.
Each column begins with $-\infty$ and ends with $+\infty$.
Otherwise, the column copies the column for $l_i$ within the \emph{zone}
of $v_i$,   where the zone of variable $v_i$ consists of rows $3i-2, 3i-1, 3i$ in
the table.  The construction is such that no matter how columns are
reordered, a lexicographic sort can rearrange rows only within their
zones.
Similarly,  the column copies the columns for $l_j$ and
$l_k$ within the zones of $v_j$ and $v_k$, respectively.  Otherwise,
the part of the column that is in the zone of $w_l$ ($l \not \in \{i,j,k\}$),
contains $0_{w_l}$.
See Fig.~\ref{fig:COMLR} for the table constructed for
$\{ 
\{ v_1, \bar{v}_2, v_3 \},
\{ \bar{v_1}, \bar{v}_2, v_3 \},
\{ \bar{v_1}, \bar{v}_3, v_4 \},
\{ v_1, v_3, v_4 \}
\}$.
Finally, we set the maximum-runs-per-column bound $K=|V|+7$.

The construction creates literal columns that cannot have many runs
no matter how we reorder columns and lexicographically sort the rows.
Consequently these columns always meet the $|V|+7$ bound. 
For clause columns: after \emph{any} column permutation 
and lexicographic sorting, a clause column
can have at most $|V|+8$ runs:
\begin{itemize} 
\item 2 for the $-\infty$ and 
the $+\infty$,
\item  $(|V|-3)$ for the variables that are not 
in the clause, 
\item and at most 3 for each of the 3 variables that are
in the clause.
\end{itemize}

\begin{figure}
\begin{centering}
$$
\begin{array}{cccccccccccc}
\bar{v}_1 & v_1       & \bar{v}_2 & v_2       & \bar{v}_3 &  v_3      &  \bar{v}_4&  v_4      &  c_1      &   c_2     &  c_3      &   c_4   \\
          &           &           &           &           &           &           &           &           &           &           &         \\
-\infty   & -\infty   & -\infty   & -\infty   & -\infty   & -\infty   & -\infty   & -\infty   & -\infty   & -\infty   & -\infty   & -\infty \\
\bar{w_1} & w_1       & -\infty   & -\infty   & -\infty   & -\infty   & -\infty   & -\infty   &  w_1      & \bar{w_1} & \bar{w_1} & w_1     \\
\bar{w_1} & 0_{w_1}   & -\infty   & -\infty   & -\infty   & -\infty   & -\infty   & -\infty   & 0_{w_1}   & \bar{w_1} & \bar{w_1} & 0_{w_1} \\
0_{w_1}   & w_1       & -\infty   & -\infty   & -\infty   & -\infty   & -\infty   & -\infty   & w_1       & 0_{w_1}   & 0_{w_1}   & w_1     \\
+\infty   &+\infty    & \bar{w_2} & w_2       & -\infty   & -\infty   & -\infty   & -\infty   & \bar{w_2} & \bar{w_2} &   0_{w_2} & 0_{w_2} \\
+\infty   &+\infty    & \bar{w_2} & 0_{w_2}   & -\infty   & -\infty   & -\infty   & -\infty   & \bar{w_2} & \bar{w_2} &   0_{w_2} & 0_{w_2} \\
+\infty   &+\infty    & 0_{w_2}   & w_2       & -\infty   & -\infty   & -\infty   & -\infty   & 0_{w_2}   & 0_{w_2}   & 0_{w_2}   & 0_{w_2} \\
+\infty   &+\infty    &+\infty    &+\infty    &\bar{w_3}  & w_3       & -\infty   & -\infty   & w_3       & w_3       & \bar{w_3} & w_3     \\
+\infty   & +\infty   & +\infty   & +\infty   & \bar{w_3} & 0_{w_3}   & -\infty   & -\infty   & 0_{w_3}   &   0_{w_3} & \bar{w_3} & 0_{w_3} \\
+\infty   & +\infty   &+\infty    & +\infty   & 0_{w_3}   & w_3       & -\infty   & -\infty   & w_3       & w_3       & 0_{w_3}   & w_3     \\
+\infty   & +\infty   & +\infty   & +\infty   & +\infty   & +\infty   & \bar{w_4} & w_4       & 0_{w_4}   & 0_{w_4}   & w_4       & w_4     \\
+\infty   & +\infty   & +\infty   & +\infty   & +\infty   & +\infty   & \bar{w_4} & 0_{w_4}   & 0_{w_4}   &   0_{w_4} &   0_{w_4} & 0_{w_4} \\
+\infty   & +\infty   & +\infty   & +\infty   & +\infty   & +\infty   & 0_{w_4}   & w_4       & 0_{w_4}   & 0_{w_4}   & w_4       & w_4     \\
+\infty   & +\infty   & +\infty   & +\infty   & +\infty   & +\infty   & +\infty   & +\infty   & +\infty   & +\infty   & +\infty   & +\infty 
\end{array}
$$

 \caption{Example construction for $\{ c_1, c_2, c_3, c_4 \},$ where 
$ c_1 = \{ v_1, \bar{v}_2, v_3 \}$,
$ c_2 = \{ \bar{v_1}, \bar{v}_2, v_3 \}$,
$ c_3 = \{ \bar{v_1}, \bar{v}_3, v_4 \}$, and
$ c_4 = \{ v_1, v_3, v_4 \}.$
\label{fig:COMLR}}
\end{centering}
\end{figure}

Table $T$ can have its columns reordered to have at most $|V|+7$
runs per column (after lexicographic sorting), if and only if the given instance of
3SAT is satisfiable.

Suppose we have a satisfying truth assignment. If $v_i$ is true,
permute the columns for $\bar{v}_i$ and $v_i$.  (Otherwise, leave
them alone.)  After permuting these columns, 
lexicographic sorting would swap the bottom two rows in the zone
for $v_i$.  Any clause containing $v_i$ would find that
this swap merges two runs of $w_i$ in its column, 
and thus we would  meet the $|V|+7$ bound 
for that clause's column.
Likewise, if $v_i$ is false, leave the two columns in their original
relationship.  The table as constructed was lexicographically sorted,
and  any clause containing $\bar{v}_i$ would continue to
have a run of $\bar{w}_i$'s and meet the run bound.  Since we have a
satisfying truth assignment, every clause column will contain at
least one such run.

Conversely, suppose we have permuted table columns such that
the lexicographically sorted table has no column with more than $|V|+7$ runs.
Because lexicographic sorting is restricted to rearranging rows only within
their zones, a clause's column must contain a length-two
run  of $w_i$ or $\bar{w}_i$, for some $1 \leq i \leq |V|$.
The construction guarantees that if any clause column contains a length-two
run of $w_i$, then no column contains a length-two run of $\bar{w}_i$.
Similarly, a length-two run of $\bar{w}_i$ precludes a length-two
run of $w_i$.  Moreover, by construction we see that a column containing
the length-two run of $w_i$ must contain $v_i$.  Hence, we set $v_i$ to true.
Likewise, for any run of $\bar{w}_i$ we set $v_i$ to false.
Clearly, this truth setting satisfies the original 3SAT instance.
\end{proof}

\end{document}